\documentclass[fleqn,usenatbib]{mnras}

\usepackage[T1]{fontenc}
\usepackage{ae,aecompl}
\usepackage[utf8]{inputenc}
\usepackage{float}
\usepackage{amsmath,amssymb}
\usepackage{mathrsfs}
\usepackage{theorem}
\usepackage{graphicx}
\usepackage{color}
\usepackage{wasysym}
\usepackage{hyperref} 
\usepackage{verbatim}

\definecolor{blue}{rgb}{0,0,1}
\definecolor{green}{rgb}{0,0.65,0.5}
\definecolor{verde}{rgb}{0.,.5,0.4}
\definecolor{marron}{rgb}{0.7,0.2,0.1}
\definecolor{red}{rgb}{1,0,0}
\definecolor{vio}{rgb}{1,0,1}
\definecolor{ama}{rgb}{1,1,0}

\title[Strong gravitational lens image of the M87...]
{\bf Strong gravitational lens image of the M87 black hole with a simple accreting 
	matter model}

\author[E.F.Boero and O.M.Moreschi]{
	Ezequiel F. Boero,$^{1}$\thanks{E-mail: ezequiel.boero@unc.edu.ar} 
	and
	Osvaldo M. Moreschi,$^{2,3}$\thanks{E-mail: o.moreschi@unc.edu.ar}
	\\
	{\rm \small $^{1}$ Instituto de Astronom\'\i{}a Te\'{o}rica y Experimental (IATE), CONICET,} \\
	{\rm \small Observatorio Astron\'{o}mico de Córdoba,}\\
	{\rm \small Laprida 854, (X5000BGR) C\'{o}rdoba, Argentina.}\\
	{\rm \small $^{2}$Facultad de Matem\'{a}tica, Astronom\'\i{}a, F\'\i{}sica y Computaci\'{o}n (FaMAF),}\\
	{\rm \small Universidad Nacional de C\'{o}rdoba,} \\ 
	{\rm \small $^{3}$Instituto de F\'\i{}sica Enrique Gaviola, IFEG, CONICET, } \\
	{\rm \small Ciudad Universitaria, (5000) C\'{o}rdoba, Argentina.} 
}

\date{Accepted XXX. Received YYY; in original form ZZZ}

\begin{document}
	\label{firstpage}
	\pagerange{\pageref{firstpage}--\pageref{lastpage}}
	\maketitle
	
\begin{abstract}
We study simulated images generated from an accretion disk surrounding
the supermassive black hole hosted in the nearby galaxy M87.
We approach the problem employing very simple accreting models inspired from
magnetohydrodynamical simulations and introducing a new recipe for dealing with the combined 
integration of the geodesic and geodesic deviation equations in Kerr spacetime,
which allows for a convenient and efficient way to manage the system of equations.
The geometry of the basic emission model is given by a two temperature thin disk in the equatorial 
plane of the black hole supplemented by an asymmetric bar structure.
We show that this configuration permits to generate the most salient features appearing
in the EHT Collaboration images of M87
with impressive fidelity.

\end{abstract}


\section{Introduction}\label{sec:Introduction}

In April 2019, the Event Horizon Telescope Collaboration (EHT) reported the first 
image ever constructed of a supermassive black hole (SMBH) 
in scales of the event horizon size using very long baseline 
interferometry (VLBI) at an observing
wavelength of 1.3mm.
As it is well know, the target object was the center of the giant elliptical galaxy M87 
hosting the emission from the compact radio source M87*. 
This system belongs to the class of low luminosity active galactic nuclei (LLAGN's)
\citep{Ho:1999ss}, commonly associated to models of hot accretion flows
\citep{Shapiro:1976fr, Ichimaru:1977uf, Yuan:2014gma}; where the accreting material is 
usually described in terms of a plasma in which ions and electrons give origin 
to a fluid that has two different temperatures. 
Then, one would not expect to observe a blackbody spectrum.
In fact, the spectral energy distribution of such systems presents features that are 
thought to be associated with emission from an \emph{optically thin and geometrically thick 
accretion disk} \citep{Akiyama:2019cqa} that could be ascribed to synchrotron 
radiation \citep{Ho:1999ss} with an observed brightness temperature in radio wavelengths 
in the range of $10^{9}-10^{10}$K. 

\begin{figure}
	\centering
	\includegraphics[clip,width=0.47\textwidth]{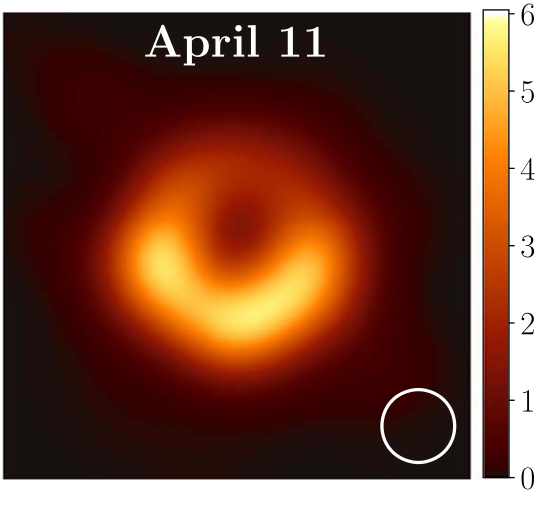}
	\caption{Reconstructed image of the innermost emitting region of M87 reported
		by the EHT Collaboration corresponding to observations during April 11 of 2017.
		This image is the result of the average of the three 
		fiducial images built with three pipelines employed by them as described in
		reference Akiyama K. et al., 2019d, ApJ, 875, L4.
		The above figure is a reproduction taken from
		fig. 15 of the same reference.
	}
	\label{fig:EHT+April_11}
\end{figure}

As shown by the EHT observations in the case of M87 (see for instance Fig. \ref{fig:EHT+April_11}), 
the emission collected at 
sub-millimeter wavelength of 1.3mm ($\sim$ 230GHz) is expected to be weakly absorbed by the 
surrounding media, allowing to view the immediate vicinity of the event horizon of the associated
supermassive black hole (SMBH) through  image reconstruction pipelines \citep{Akiyama:2019bqs}.
 
The hypothesis of a SMBH is the most ubiquitous one to which ascribe the geometry 
that allows the very energetic features observed such as the impressive jets in AGN's
\citep{BlandfordZnajek_MNRAS_1977}.

The strong gravitational field regime of the SMBH within a few 
radius in units of the mass of the black hole should then, 
imposes a relevant influence on
the propagation of radiation coming from the innermost regions. 
It is then of great importance, to study the role played by the geometry and the emission 
properties of the plasma on the images that one would observe with instruments such 
as the EHT facilities.
In particular, simulated images of accurate and plausible models need to properly give account 
of all those possible effects in order to get a clear and confident interpretation of 
the available data.
Most of the numerical works designed to achieve this goals, usually combine highly sophisticated
general relativistic magnetohydrodynamical(GRMHD) simulations together with ray tracing algorithms 
in order to produce high resolution images\citep{Noble_2007CQGra}. 
This approach requires to manage the integration of the geodesic equations 
\eqref{eq:ell+nabla+ell} with a sufficient large number of rays. 
It is for this reason that, to address these concerns, some authors have 
made use of GPU parallel computing methods; for example in \cite{Chan:2017igo}. 
We here present another approach which makes use of simplified emission models and a very 
efficient\citep{Boero:2019zkq} calculation that does not need for GPU technology in order to 
build images that can reproduce the main features of the EHT Fig. \ref{fig:EHT+April_11} 
with high fidelity.
In fact, our model provides
a realistic realization of the 
image at relatively low computational cost, with runs that last
for few minutes in a laptop.
We consider an accretion disk on a Kerr background consisting of a rotating geometrically 
thin disk with two temperature regions to which we ascribe the main emission of the plasma 
surrounding the immediate vicinity of the SMBH together with 
a supplementary asymmetric bar-like structure. 
Contributions from the geometrically thick disk then will be neglected 
in the generation of the images.
In particular, the absorption along null geodesics will be considered negligible 
in the thick disk, 
so that this region will be taken as optically thin.
For the sake of clarity, it is also important to stress that although the accretion disk 
is assumed to contribute in a significant way to the amount of observed emission and its 
characteristic, its contribution to the metric of the spacetime will be here considered
negligible. This is usually justified in most GRMHD 
treatments\citep{Abramowicz:2011xu,Font:2008fka}.
Therefore, our calculations are based on the integration of
the past directed null geodesics in the Kerr geometry starting at the observer location
together with the integration of the geodesic deviation equations,
until we reach the emitting regions.
We have explored several geometrical settings until we arrived
at the best configuration that can reproduce the April 11 EHT image.
We will show in our results that such assumption can provide an excellent description of
the bright crescent sector that encompasses a central darker
region together with the observed substructure 
at the east and south-west locations in position angle (PA) (Fig. \ref{fig:EHT+April_11}).
We present the discussion of our simulated images of M87 by considering 
different choices of the angular momentum parameter,
that were used to select a model that could accurately represent the main features
of the observed images.

In this work we will also be focusing on the lensing technique employed to carry out 
ray-tracing and the image generation.
The additional difficulties that would arise adding the 
\emph{geodesic deviation equations},
in comparison with only tracking the photons path through ray tracing on individual null geodesics,
is overcome thanks to recent new expressions for the curvature scalar sourcing the 
\emph{geodesic deviation equations} in Kerr spacetime\citep{Boero:2019zkq}. 
We present a method that innovates over previous ones often
found in the literature, in the sense that it allows for a 
joint efficient integration of the
\emph{geodesic equations} and \emph{geodesic deviation equations} with high precision;
introducing a remarkable simplifications at the moment 
of dealing with thin bundles
of null geodesics, synthesizing in a short formula previous 
results\citep{Pineault1977ApJ,James:2015yla}
obtained by brute force calculations.
More precisely, in reference \cite{James:2015yla} they employ 
from page 29 to page 31 to present the Weyl component $\Psi_{0}$.
Instead our expression \eqref{eq:Psi0} below consists of one single
term of simple evaluation.
In reference \cite{Boero:2019zkq}, we have employed those new results 
to asses the accuracy of the usual expressions for the gravitational optical scalars in the weak field 
regime\citep{Gallo11,Boero:2016nrd} 
with respect to the exact ones; in this article
instead, we employ them to generate reliable images of the innermost regions of a SMBH taking 
as precedent the recently reported observations of M87.
The approach that we propose here differs in some aspects from the usual ray tracing 
method\citep{Cunningham_1973ApJ,Akiyama:2019bqs,Nalewajko:2019mxh} 
namely, the usage of the geodesic deviation equations allows us to asses the 
contribution of the magnification effect due to local distortions of thin bundles of null 
geodesics.
This means that we can take into account the gravitational effects on the
magnitude of the observed fluxes coming from the emitting models.
Our approach also intends to be a contribution to the techniques 
that seek to achieve great quality 
of images,
such as those mentioned in \cite{James:2015yla}. 

We also find that, 
in agreement with previous studies on the subject\citep{Akiyama:2019cqa} the 
asymmetric shape of the observed crescent sector is mainly associated to the inclination 
of the spin and 
its magnitude; in particular this asymmetry in brightness is a robust indicator 
of the orientation of the 
spin axis. 
From several calculations using different orientations of the black holes spin,
we find that the favored spin direction is opposite to the observed jet,
in agreement with the EHT findings.
We show that even though the photon paths are highly bent 
in the strong field region, 
the role of the magnification is not as important,
in such region, as the enhancement
of the flux produced by the Doppler/gravity-shift of the accreting fluid.

The paper is organized as follows:
In section \ref{sec:Exact+grav+lens+opt+scalars} we present our approach based in the notion
of exact gravitational lens optical scalar which allows for the joint implementation of the 
geodesic and geodesic deviation equations.
In section \ref{sec:Null+geod+Kerr} we apply the equations 
to the specific case of Kerr geometry.
Section \ref{sec:accretion+disk} contains our simplified model for the emission of the 
accretion disk.
In section \ref{sec:Numerical+implementaion} we describe the details of our numerical 
implementation and in section \ref{sec:Results} we present our images and further results.
The last section concludes with some final comments.

Whenever convenient along the text, we will employ standard abstract index notation 
with Latin letters $a,b,c,...$ for tensor fields. 
Our choice of signature for the spacetime metric $g_{ab}$ is $(+,-,-,-)$.

\section{Exact gravitational lens optical scalars}\label{sec:Exact+grav+lens+opt+scalars}
The physical situation we want to address is that regarding the generation of 
images due to a classical observation into our past light cone.

We will make the convenient choice to deal with past directed null geodesics leaving 
the observer and reaching the sources.
On the past light cone of the observer we set a coordinate system in the 
following way: two angular coordinates $(\theta, \phi)$ label each one of the 
null geodesics 
generating the past null cone an the third one is taken to be the affine parameter 
$\lambda$ that grows into the past and has a scale set by the normalizing condition at the apex 
of the cone:
\begin{equation}\label{eq:ell+v+minus1}
\ell^a v_a = -1,
\end{equation}
where $v^a$ denotes the 4-velocity vector of the observer and $\ell^a$ is the tangent vector 
to a null geodesic with affine parameterization, i.e. $\ell^a$ satisfies
\begin{equation}\label{eq:ell+nabla+ell}
\ell^a \nabla_a \ell^b = 0.
\end{equation}

We also consider a null tetrad $(\ell^a, m^a ,\bar{m}^a, n^a)$ where $m^a$ 
and $\bar{m}^a$ is a pair of complex conjugated vectors and $n^a$ an additional 
real null vector.
It will be important to choose the two complex vectors $m^a$ and $\bar{m}^a$ to be parallel 
propagated along the geodesic since they ensure the simplicity in our specific treatment 
on the Kerr geometry as we will mention later in section \ref{sec:Null+geod+Kerr}.

Along the path of the photons, one also considers thin bundles of null geodesics; 
where
local distortions of such bundles give origin to changes in the observed intensity 
with respect to the `unlensed' situation in which the propagation would take place on a spacetime with 
no curvature.
Then, in order to describe such changes one introduces the geodesic deviation vector 
$\varsigma^a$, that in terms of our tetrad can be written in the following way:
\begin{equation}\label{eq:varsigma}
\varsigma^a = \varsigma \bar{m}^a + \bar{\varsigma} m^a + \eta \ell^a;
\end{equation}
and by definition \citep{Wald84}, $\varsigma^a$ must be Lie transported, this is:
\begin{equation}\label{eq:Lie+prop+devVec}
\mathscr{L}_{\ell} \; \varsigma^{a} = 0.
\end{equation}
The component $\eta$ measures a difference of position along the
null geodesic, and therefore it is not related to neighboring geodesics
and it does not play any further physical role.
This is the reason sometimes this quantity is completely neglected.

The complete description of thin bundles of null geodesics is accomplished by means of 
the coupled system of geodesics equations \eqref{eq:ell+nabla+ell} and the geodesic deviation equations:
\begin{equation}\label{eq:geod+dev+equation}
\ell^a \nabla_a \left( \ell^b \nabla_b \varsigma^d \right) = 
R_{abc}^{\;\;\;\;\;d}\ell^a \varsigma^b \ell^c;	
\end{equation}
where as usual here, $R_{abc}^{\;\;\;\;\;d}$ denotes the Riemann tensor associated to the 
Levi-Civita connection $\nabla_a$.

\subsection{Optical scalars}

Let us introduce in this section the notion of \emph{exact gravitational lens 
optical scalars} which involves the exact integration of the coupled 
system of equations \eqref{eq:ell+nabla+ell} and \eqref{eq:geod+dev+equation} 
above mentioned.

The common notion of optical scalars is meanly present in weak lensing studies
and comes from the comparison between the observations of photons propagating 
in the physical curved spacetime and the hypothetical situation where no 
lensing takes place (i.e.: no gravity, and so flat geometry).
However, one can extrapolate this concept to more general situations where 
not necessarily the photons travel across the weak field zone of the 
spacetime; below we explicitly show how to obtain this generalization.

Each pixel of the images that is detected by an optical device coming from the 
direction $\theta^a$ on the sky and corresponding to the null tangent vector 
$\ell^a$, can be associated to a thin bundle of null geodesics subtending an angle 
$\delta \theta^a$.
In absence of curvature the same portion of the source would be observed 
in another direction, let us say $\beta^a$ which would have associated a null 
vector $\ell'^a$ and subtending an angle $\delta \beta^a$. 

The optical scalars $\kappa$, $\gamma_{1}$, $\gamma_{2}$ and $\omega$ are defined 
through the linear relation between $\delta \theta^a$ and 
$\delta \beta^a$
\begin{equation}\label{eq:linear+relation+angles}
\delta \beta^a = \mathcal{A}^a_{\;\,b} \delta \theta^b,
\end{equation}
with the matrix $\mathcal{A}^a_{\;\,b}$ 
\begin{equation}
\mathcal{A}^a_{\;\,b} = \begin{pmatrix}
1 - \kappa - \gamma_1 & -\gamma_2 - \omega \\
-\gamma_2 + \omega    & 1 - \kappa + \gamma_1
\end{pmatrix}.
\end{equation}

For a given angular aperture, the relation between $\delta \theta^a$
and the deviation vector $\varsigma^a$ is given by
\begin{equation}
\delta \theta^a \equiv 
\left. 
\begin{pmatrix}
\upsilon_{R}^{} \\
\upsilon_{I}^{}
\end{pmatrix}^a
\right|_{\lambda=0}
\equiv 
\left.
\begin{pmatrix}
\ell \left(\varsigma_{R}^{}\right) \\
\ell\left( \varsigma_{I}^{} \right)
\end{pmatrix}^a
\right|_{\lambda=0}
,
\end{equation}
where the right hand side is evaluated at the observer position and
where 
\begin{equation}
\left. \varsigma^a \right|_{\lambda = 0} 
= 
\left.
\begin{pmatrix}
\varsigma_{R}^{} \\
\varsigma_{I}^{}
\end{pmatrix}^a
\right|_{\lambda = 0}
= 
\begin{pmatrix}
0 \\
0
\end{pmatrix}^a
,
\end{equation}
and, $\varsigma_{R}^{}$ and $\varsigma_{I}^{}$ are the real and imaginary components
of the deviation vector $\varsigma^a$.
The integration of equations \eqref{eq:ell+nabla+ell} and \eqref{eq:geod+dev+equation} 
from $\lambda = 0$ to $\lambda = \lambda_{s}$ allows to obtain the fiduciary angle
$\delta \beta^a$ as follows:
\begin{equation}
\delta \beta^a \equiv \lambda_s 
\begin{pmatrix}
\varsigma_{R}^{} \\
\varsigma_{I}^{}
\end{pmatrix}^a
,
\end{equation}
where $\lambda_s \equiv \lambda_\text{observer} -  \lambda_\text{source}$.
Equation \eqref{eq:linear+relation+angles}, then explicitly becomes:
\begin{align}
\varsigma_{R}^{} =& \big( 1 - \kappa - \gamma_1 \big) \upsilon_{R}^{}  - 
\big(\gamma_2 + \omega \big) \upsilon_{I}^{}, \\
\varsigma_{I}^{} =& - \big(\gamma_2 - \omega \big) \upsilon_{R}^{} 
+ \big(1 - \kappa + \gamma_1 \big) \upsilon_{I}^{}. 
\end{align}

Then, the optical scalars are found simply by considering a pair of mutually orthogonal deviation 
vectors, $\varsigma_1^a$ and $\varsigma_2^a$ with initial conditions at the observer 
position:
\begin{align}
\left. 
\begin{pmatrix}
\varsigma_{R_1}^{} \\
\varsigma_{I_1}^{}
\end{pmatrix}^a
\right|_{\lambda=0}
=&
\begin{pmatrix}
0 \\
0
\end{pmatrix}^a
,
\\
\left. 
\begin{pmatrix}
\upsilon_{R_1}^{} \\
\upsilon_{I_1}^{}
\end{pmatrix}^a
\right|_{\lambda=0}
=&
\begin{pmatrix}
1 \\
0
\end{pmatrix}^a
,
\end{align}
and
\begin{align}
\left. 
\begin{pmatrix}
\varsigma_{R_2}^{} \\
\varsigma_{I_2}^{}
\end{pmatrix}^a
\right|_{\lambda=0}
=&
\begin{pmatrix}
0 \\
0
\end{pmatrix}^a
,
\\
\left. 
\begin{pmatrix}
\upsilon_{R_2}^{} \\
\upsilon_{I_2}^{}
\end{pmatrix}^a
\right|_{\lambda=0}
=&
\begin{pmatrix}
0 \\
1
\end{pmatrix}^a
,
\end{align}
that yields the final linear system of equations for the quantities
$\kappa$, $\gamma_{1}$, $\gamma_{2}$ and $\omega$:
\begin{align}
\varsigma_{R1}^{} =& 
\big( 1 - \kappa - \gamma_1 \big) \lambda_s , \\
\varsigma_{I1}^{} =& - \big(\gamma_2 - \omega \big)\lambda_s , \\
\varsigma_{R2}^{} =& -\big(\gamma_2 + \omega \big)\lambda_s , \\
\varsigma_{I2}^{} =& \big( 1 - \kappa + \gamma_1 \big)\lambda_s,
\end{align}
or equivalently 
\begin{align}
\kappa =& 1 - \frac{\varsigma_{R1}^{} + \varsigma_{I2}^{}}{2 \lambda_s} , \\
\gamma_1 =& \frac{\varsigma_{I2}^{} - \varsigma_{R1}^{}}{2 \lambda_s}, \\
\gamma_2 =& - \frac{\varsigma_{I1}^{} + \varsigma_{R2}^{}}{2 \lambda_s}, \\
\omega =& \frac{\varsigma_{I1}^{} - \varsigma_{R2}^{}}{2 \lambda_s}.
\end{align}

Of crucial importance in our discussion of the observed flux is the 
magnification factor $\mu$ which is given in terms of the above quantities as follows:
\begin{equation}\label{eq:magnification}
\mu = \frac{1}{\left(1 - \kappa \right)^2 - \left(\gamma_1^2 + \gamma_2^2 \right) + \omega^2}
=
\frac{\lambda_s}{ \varsigma_{I2}^{} \varsigma_{R1}^{} - \varsigma_{I1}^{} \varsigma_{R2}^{}}.
\end{equation}	

The above prescription will be then implemented in our numerical simulations to compute and 
asses the magnification of our constructed images of section \ref{sec:Results}.
	
\section{Null geodesics and the null geodesic deviation equation in Kerr spacetime}
\label{sec:Null+geod+Kerr}

\subsection{Kerr metric}\label{subsec:Kerr+metric} 
Here we adapt the discussion of the preceding section to the particular case of 
Kerr metric. 
We will use Boyer-Lindquist coordinate\citep{Boyer67} system $(t,r,\theta, \phi)$ 
in which the line element is:
\begin{equation}\label{eq:Kerr-usualBL-form}
\begin{split}
ds^2 
=& \left( 1 - \varPhi\right) dt^2  
+ 2 \varPhi a \sin(\theta)^2 dt d\phi 
- \frac{\Sigma}{\Delta} dr^2 
\\
& - \Sigma d\theta^2 
- \left( r^2 + a^2 + \varPhi a^2  \sin^2(\theta)\right) \sin(\theta)^2
d\phi^2
;
\end{split}
\end{equation}
where $M$ and $a$ are the mass and rotation parameter respectively, and the functions 
$\Sigma(r, \theta) $, $\Delta(r)$ and $\varPhi(r,\theta)$, are 
defined as
\begin{equation}\label{eq:metric+func+Sigma}
\Sigma = r^2 + a^2 \cos(\theta)^2 , 
\end{equation}
\begin{equation}\label{eq:eq:metric+func+Delta}
\Delta = r^2 - 2rM + a^2 ,  
\end{equation}
\begin{equation}\label{eq:eq:metric+func+Phi}
\varPhi =\frac{2 M r}{\Sigma} .
\end{equation}

In these coordinates the \emph{outer event horizon} is given 
implicitly by one of the solutions to the condition
$\Delta(r) = 0$, namely $r_+$: 
\begin{equation}\label{eq:r+horizon}
r_+ = M + \sqrt{M^2 - a^2}.
\end{equation}

\subsection{Null geodesics}\label{subsec:Null+geod+Kerr}

The discussion of null geodesic motion in Kerr spacetime is central to the 
image generation, so we will summarize the essential and necessary
equations for this work below.
It is well known that Kerr geometry has the quite remarkable property 
that geodesics admit a complete set of first integrals of motion\citep{Carter:1968rr, Chandrasekhar:1983este}: 
the energy $E$, the angular momentum $L$, the Carter's constant $K$ and, the last one is
$g_{ab} \ell^a \ell^b = 0$ which ensures the null character.

These conserved quantities have the following explicit expressions
\begin{equation}\label{eq:E+geod}
E \equiv  g_{ab}\ell^a \xi_t^a 
=  \left(1 - \Phi \right)\dot{t} + a\Phi \sin(\theta)^2 \dot{\phi},
\end{equation}
\begin{equation}\label{eq:Lz+geod}
\begin{split}
L \equiv& -g_{ab}\ell^a \xi_\phi^b \\
=& - a\Phi \sin(\theta)^2 \dot{t} 
+
\left( r^2 + a^2 + a^2\Phi \sin(\theta)^2 \right)\sin(\theta)^2\dot{\phi} ,
\end{split}
\end{equation}
\begin{equation}\label{eq:K+Carter+geod}
\begin{split}
K \equiv& 2 \Sigma \ell^a \ell^b \tilde{\ell}_a \tilde{n}_b \\
=& \Delta \left( \dot{t} - \frac{\Sigma \dot{r}}{\Delta} - a \sin(\theta)^2 \dot{\phi} \right)
\left( \dot{t} + \frac{\Sigma \dot{r}}{\Delta} - a \sin(\theta)^2 \dot{\phi} \right),
\end{split}
\end{equation}
where $\ell^a = \left( \dot{t}, \dot{r}, \dot{\theta}, \dot{\phi}\right)$ and
$\xi_t^a = \partial_t^a$ , $\xi_\phi^a = \partial_\phi^a$ are the time-translation
and axial Killing vectors respectively; while the null condition is
\begin{equation}\label{eq:null+condition+geod} 
\begin{split}
0 
=& \left( 1 - \varPhi\right) \dot{t}^2  + 2 \varPhi a \sin(\theta)^2 \dot{t} \dot{\phi} 
- \frac{\Sigma}{\Delta} \dot{r}^2 
\\
& - \Sigma \dot{\theta}^2 - 
\left( r^2 + a^2 + \varPhi a^2  \sin(\theta)^2 \right) \sin(\theta)^2
\dot{\phi}^2
\\
=&
\big[ \left(1 - \Phi \right)\dot{t} + a\Phi \sin(\theta)^2 \dot{\phi} \big] \dot{t}
- \frac{\Sigma}{\Delta} \dot{r}^2 - \Sigma \dot{\theta}^2
\\
& + 
\big[ a\Phi \sin(\theta)^2 \dot{t} -
\left( r^2 + a^2 + a^2\Phi \sin(\theta)^2 \right)\sin(\theta)^2\dot{\phi} \big]\dot{\phi}
\\
=&
E \dot{t} - \frac{\Sigma}{\Delta} \dot{r}^2 - \Sigma \dot{\theta}^2 - L \dot{\phi}
.
\end{split}
\end{equation}
Let us note that these definitions are consistent with the usual convention in which, 
$E$ is negative while that $L$ is positive for direct (or prograde) past directed null geodesics. 

The above system of equations is usually written in a more convenient form\citep{Chandrasekhar:1983este}:
\begin{align}
\Sigma \dot t 
=& 
\frac{1}{\Delta}\left[ E \Big(
(r^2 + a^2)^2 - \Delta \, a^2 \sin(\theta)^2 \Big)
- 2 a M r L \right]
, 
\label{eq:ellt-geod}
\\
\Sigma^2 \dot r^2 =& \mathcal{R}(r)
, 
\label{eq:ellr-geod}
\\
\Sigma^2 \dot \theta^2 =& \varTheta(\theta) 
,
\label{eq:elltheta-geod}
\\
\Sigma \dot \phi =& 
\frac{1}{\Delta}\left[ 2E a M r + (\Sigma - 2Mr) \frac{L}{\sin(\theta)^2}\right]
;
\label{eq:ellphi-geod}
\end{align}
where the functions $\mathcal{R}(r)$ and $\varTheta(\theta)$ are defined as
\begin{align}
\mathcal{R}(r) =& \Big( E\left(r^2 + a^2 \right) - a L \Big)^2 - K \Delta(r) 
,
\label{eq:R+geod+func}
\\
\varTheta(\theta) =& K - \left( \frac{L}{\sin(\theta)} - a E \sin(\theta) \right)^2 
.
\end{align}

Let us note that for an stationary observer 
the constant $E$ only depends on the position through:
\begin{equation}
E = -\sqrt{1 - \varPhi_o};
\end{equation}
here the subindex $_o$ denotes evaluation at the observer position.
This is just a consequence of equation \eqref{eq:ell+v+minus1}.

The other constants	 of motion determine the angular direction of the incoming photons 
in the sky of the observer. 
In this respect it is important to mention that the choice of a suitable frame of reference 
is a subtle question that one has to address due to non-static nature of the geometry; 
for instance, in the typical situation of an observer far from the central region;
the coordinate frame works fairly well for all practical purposes.
For more general purposes, a suitable frame of reference that captures the notion of center 
of the BH was introduced in 
\cite{Boero:2019zkq}
making use of the null congruence 
belonging to center of mass \citep{Arganaraz:2021fpm} passing by the observer position.
In the present case, both settings agree with high accuracy since our observations 
take place very 
far from the central region of the nearest SMBH. 
One can then corroborate that the correspondence among $(L, K)$ and the observational angular 
coordinates $\left( \alpha_x, \delta_z \right)$ is as follows:
\begin{align}
\alpha_{x}
=&
\frac{L_z}{r_o \sin(\theta_o)}
, 
\\
\begin{split}
\delta_{z} 
=&
\frac{(\pm)}{r_o} 
\left[K - \left( \frac{L_z}{\sin(\theta_o)} - aE\sin(\theta_o) \right)^2\right]^{1/2}
\\
=&
\frac{(\pm)}{r_o} 
\left[K 
- \left( \sqrt{K_{o_\text{cm}}} - \frac{L_z}{\sin(\theta_o)}  \right)^2
\right]^{1/2}
.
\end{split}
\end{align} 
where $K_{o_{\text{cm}}}$ is the value of the Carter's constant of the central null 
geodesic of the center of mass and is given as a limit by the expression below:
\begin{equation}
K_{o_{\text{cm}}} = E^2 a^2 \sin(\theta_o)^2.
\end{equation}

\subsection{Null geodesic deviation equation}\label{subsec:dev+eq+Kerr}
The aim of this brief subsection is to present a compact and useful form of the 
geodesic deviation equation valid for Kerr spacetime. 
In the coordinate system adapted to the pass null cone of the observer,
the geodesic deviation equation \eqref{eq:geod+dev+equation} can be conveniently 
written as a first order system in terms of the tuple 
$\left(\varsigma, v_{\varsigma}, \bar{\varsigma}, \bar{v}_{\varsigma} \right)$
as follows: 
\begin{align}
\dot{\varsigma} =& v_{\varsigma}, \\
\dot{v}_{\varsigma} =&  - \Psi_0 \bar{\varsigma}, \\
\dot{\bar{\varsigma}} =& \bar{v}_{\varsigma}, \\
\dot{\bar{v}}_{\varsigma} =& - \bar{\Psi}_0 \varsigma,
\end{align}
where the Weyl curvature scalar $\Psi_{0}$\citep{Geroch73} has the 
expression\citep{Boero:2019zkq}: 
\begin{equation}\label{eq:Psi0}
\Psi_0 = - \frac{3 M^{5/3} \mathbb{K}^2}{2 \big(r - ia \cos(\theta) \big)^5}; 
\end{equation}
the constant quantity $\mathbb{K}$ is a spin-weight constant given by:
\begin{equation}\label{eq:K}
\begin{split}
\mathbb{K} =& 
-\frac{i \sqrt{2}}{ M^{1/3}}\left[
\delta_z r_o
- 
i\left(\sqrt{K_{o_\text{cm}}} + \alpha_x r_o \right)\right]
.
\end{split}
\end{equation}

As mentioned previously,
the above expressions \eqref{eq:Psi0} and \eqref{eq:K} 
 constitute a great improvement with respect to the lengthy formulas 
previously reported in the literature\citep{Pineault1977ApJ, Pineault1977ApJ_II} and permit 
an efficient calculation in numerical implementations.

\section{The geometry and kinematics of the accretion disk model}\label{sec:accretion+disk}

\subsection{A geometrically thin accretion region for the main emission component}
We here proceed to describe our simple model for the geometry of the accretion disk that
will try to capture the main features in the EHT images of M87.
Perhaps, it would be of help to mention that we are not concerned here with the detailed 
astrophysical modeling of the plasma surrounding the SMBH, neither with its whole dynamical 
evolution which is usually investigated through sophisticated general relativistic 
magnetohydrodynamical (GRMHD) simulations \citep{Font:2008fka}.
Our model instead, focuses on the imaging process due to a reduced amount of elements,
rather than 
to give a complex full account of process such as the jet production or the 
characteristic spectral distributions at certain wavelengths. 

Our source is considered to be a thin and opaque disk on Kerr equatorial plane, 
whose radius is only few units in terms of the mass of the black hole,
with an energy-momentum tensor of negligible influence on the fixed Kerr background
representing the SMBH.
That is,  
we will assume that the main contribution to the observed intensity 
comes from this thin disk. 
This is, in fact a feature shared by several GRMHD models, in particular for those referred
as MAD by `magnetically arrested disk' \citep{Narayan_2003PASJ}.

The origin of the emission is supposed to be of synchrotron type due to the motion of 
the electrons and ions in the hot plasma comprising the accretion disk.
This would be implicit in our description, but since we are not concerned with the specific
processes of emission and due to the fact that observations are done within a bandwidth 
$\delta \nu_o$ given by the instrumental arrangement, it would be enough for the present 
purposes to speak of the observed intensity $I(\nu_o)$.
In this regards, it is also conventional to make reference of the observed intensity through 
the so called brightness temperature, $T_o$ which is defined through the Rayleight-Jeans
expression:
\begin{equation}\label{eq:R-J}
I(\nu_o,T_o) = \frac{2  \nu_o^2 k_B T_o}{c^2}.
\end{equation}
In particular, it should be noted that in reference \cite{Akiyama:2019bqs} they have chosen 
to build the graphs around the quantity $T_o$. 
In our graphs, instead we will use an arbitrary scale based on the idea of a two temperature 
thin disk, that we explain below. 

The model is inspired in that considered in reference \cite{Moscibrodzka:2015pda} 
and depicted in the central graph of fig. 2 in that article,
taking two representative values for the temperatures in the equatorial plane.
This is an important feature that is also in agreement with most models 
for LLAGNS's\citep{Yuan:2014gma}.
The innermost region having a temperature $T_{\text{in}}$ higher 
than the outermost one at $T_{\text{out}}$.
We will show later that this simple prescription provides images that resemble 
fairly well those observed by the EHT Collaboration in the case of M87; for instance, 
this is enough for obtaining most of the large crescent at the south-west location.
An enhanced modeling that gives account of the whole structure so far observed,
including the two maxima in brightness in the south crescent
shown in Fig. \ref{fig:EHT+April_11}, 
is obtained by considering a one side bar feature emitting at a temperature $T_{\text{bar}}$.
The specific values of the intervening quantities will be discussed further in
section \ref{sec:Results} where we will also discuss the results produced 
by these configurations and the influences of the spin of the geometry.

\subsection{The surface brightness and the observed specific intensity of a thin and opaque disk}
In order to relate the emission properties of the accretion disk with the 
flux $\mathscr{F}$ measured by our facilities on Earth coming from a solid angle $d\Omega_o$, 
let us recall that 
\begin{equation}
\mathscr{F} = I(\nu_o, T_o) d\Omega_o;
\end{equation}
where $I(\nu_o, T_o)$ is the \emph{specific intensity} at the observer position.
Then, if there is no absorption and the radiation propagates through an homogeneous media,
the so-called Etherington theorem\citep{Etherington33, Ellis71} 
(see also appendix \ref{ap:B+I})
relating angular 
distances in a general spacetime ensures that:
\begin{equation}\label{eq:surfBright+specifIntens}
B(\nu, T) = (1 + z)^3 \, I(\nu_o, T_o) ,
\end{equation}
where $B(\nu, T)$ is the \emph{surface brightness} of the source and 
the redshift $z$, defined by the relation
\begin{equation}\label{eq:redshift}
1 + z = \frac{\ell_a u_e^a}{\ell_a u_o^a}
= - \ell_a u_e^a;
\end{equation}
gives account of the shift between the observed and the emitted frequencies $\nu_o$ and $\nu$
respectively:
\begin{equation}
1 + z = \frac{\nu}{\nu_o}.
\end{equation}
Let us note that the minus sign in equation \eqref{eq:redshift} follows due to equation 
\eqref{eq:ell+v+minus1} in our setting.

In cases where absorption and emission along the geodesic are considered, 
equation \eqref{eq:surfBright+specifIntens} is modified in order to take account of these
contributions; arriving to the so-called \emph{radiative transfer equation} 
\citep{Spitzer_1998book, Ellis2012}.
In this work, for simplicity we neglect these effects, we just take into account  
that observations are done within a bandwidth $\delta \nu_o$ given by the instrumental 
arrangement and there is no attenuation, so that the flux per unit of solid angle is given by:
\begin{equation}\label{key}
I(\nu_o) \delta \nu_o
=
\frac{1}{(1+z)^3} B(\nu) \delta \nu_o
=
\frac{1}{(1+z)^4} B(\nu) \delta \nu
.
\end{equation}
In terms of fluxes, we have that
\begin{equation}\label{eq:flux+frequency}
\mathscr{F} \delta \nu_o = \frac{\mu}{\left(1 + z \right)^4}\mathscr{F}_0 \delta \nu_s 
;
\end{equation}
where we note the presence of the magnification defined in equation \eqref{eq:magnification}
and $\mathscr{F}_0 \equiv \mathscr{F}_0(\lambda,z)$ denotes the flux that one would expect 
to collect from the same observed object at the same distance $\lambda$ in Minkowski space–time 
with the same relative motion that is actually observed.
For more details we refer the reader to appendix \ref{ap:B+I}.

\subsection{Timelike geodesics}

Kinematical information about the motion of small pieces of the fluid on the accretion 
disk surrounding the SMBH is needed in our treatment 
in order to give account of the redshift
due to peculiar motions. 
In this subsection we present our modeling that assumes that the kinematics is 
derived from geodesic motion. 
Then, we turn now to recall some well known facts\citep{Chandrasekhar:1983este, Bardeen:1972fi} 
about timelike geodesics; the main result is the existence of a whole set of first integrals 
in a completely analogy with the null case; we have:
\begin{align}
\Sigma \dot t_e
=& 
\frac{1}{\Delta}\left[ E_e \Big(
(r_e^2 + a^2)^2 - \Delta \, a^2 \sin^2(\theta_e) \Big)
- 2 a M r_e L_e\right]
, 
\label{eq:ellt-geod+timelike}
\\
\Sigma^2 \dot r_e^2 =& \mathcal{R}_e(r_e) - r^2 \Delta(r_e)
, 
\label{eq:ellr-geod+timelike}
\\
\Sigma^2 \dot \theta_e^2 =& \varTheta_e(\theta_e) - a^2 \cos(\theta_e)^2
,
\label{eq:elltheta-geod+timelike}
\\
\Sigma \dot \phi_e =& \frac{1}{\Delta}\left[ 2E_e a M r_e + (\Sigma - 2Mr_e) \frac{L_e}{\sin^2(\theta_e)}\right]
;
\label{eq:ellphi-geod+timelike}
\end{align}
where the constants of motion $E=E_e$, $L=L_e$, $K=K_e$ referring to the `emitters'
are defined in this case as follows:
\begin{equation}\label{eq:E+geod+timelike}
E_e \equiv  g_{ab} u^a \xi_t^a 
=  \left(1 - \Phi \right)\dot{t}_e + a\Phi \sin(\theta)^2 \dot{\phi}_e,
\end{equation}
\begin{equation}\label{eq:Lz+geod+timelike}
\begin{split}
L_e \equiv - g_{ab} u^a \xi_\phi^b 
=& - a\Phi \sin(\theta_e)^2 \dot{t}_e  \\
& +
\left( r_e^2 + a^2 + a^2\Phi \right)\sin(\theta_e)^2\dot{\phi}_e ,
\end{split}
\end{equation}
\begin{equation}\label{eq:K+Carter+geod+timelike}
\begin{split}
K_e \equiv& 2 \Sigma u^a u^b \tilde{\ell}_a \tilde{n}_b + r_e^2  u^a u_a 
\\
=& \Delta \left( \dot{t}_e - \frac{\Sigma}{\Delta}\dot{r_e} - a \sin(\theta_e)^2 \dot{\phi} \right)
\left( \dot{t}_e + \frac{\Sigma}{\Delta} \dot{r}_e - a \sin(\theta_e)^2 \dot{\phi} \right)
\\
& - r_e^2,
\end{split}
\end{equation}
and,
$\mathcal{R}_e(r_e)$ and $\varTheta_e(\theta_e)$ are in turn defined as
\begin{align}
\mathcal{R}_e(r_e) =& \Big( E_e \left(r_e^2 + a^2 \right) - a L_e \Big)^2 - K_e \Delta(r_e)
,
\label{eq:R+geod+func+timelike}
\\
\varTheta_e(\theta_e) =& K_e - \left( \frac{L_e}{\sin(\theta_e)} - a E_e \sin(\theta_e) \right)^2 
.
\end{align}

\subsubsection{Circular orbits on the equatorial plane}
As a working assumption in the description of a geometrically thin disk,
we will take a fluid whose distribution is confined to the equatorial 
plane and with negligible radial stream variations, namely 
$\dot{r} \simeq 0$ in comparison to its axial motion $\dot{\phi}$.
So, here we discuss the case of circular orbits on the equatorial plane 
which is the simplest working assumption for the motion of matter in a 
thin disk surrounding the BH.
Then, an emitter with circular orbit is given by 
$(r=r_e,\theta=\pi/2)$ with $r_e$=constant; and its four velocity 
$u_e^a = \left( \dot{t}_e, 0, 0, \dot{\phi}_e \right)^a$ satisfies 
\citep{Chandrasekhar:1983este}: 
\begin{equation}\label{eq:tdot}
\begin{split}
r_e^2 \dot t_e 
=& 
\frac{1}{\Delta_e}\left[
E_e \bigg(
(r_e^2 + a^2)^2 - \Delta \, a^2 
\bigg)
- 2 a M r_e L_e\right]
,
\end{split}
\end{equation}

\begin{equation}\label{eq:rdot}
\begin{split}
0 =&  \mathcal{R}_e(r_e) - r_e^2 \Delta(r_e) 
\\ 
=& \Big( E_e (r_e^2 + a^2) - a L_e \Big)^2 - \Delta (r_e^2 + K_e)
,
\end{split}
\end{equation}
\begin{equation}\label{eq:titadot}
\begin{split} 
0 =& \varTheta_e(\pi/2)  - a^2 \cos(\pi/2)^2
\\
=& K_e - (E_e a - L_e)^2
,
\end{split}
\end{equation}
as a consequence of $(r_e^2 \dot r_e)^2 = 0$ and $(r_e^2 \dot \theta_e)^2 =  0$
respectively; and,
\begin{equation}\label{eq:fidot}
\begin{split}
r_e^2 \dot \phi_e
=& \frac{1}{\Delta}\Big[ 2 E_e a M r_e + (r_e^2 - 2 M r_e) L_e \Big]
;
\end{split}
\end{equation}
to which we must add the condition $1 = g_{ab} u^a u^b$, that is:
\begin{equation}\label{eq:ueue}
\begin{split}
1 =& 
\big[ \left( 1 - \varPhi \right) \dot t_e  + a \varPhi \dot\phi_e \big] 
\dot t_e
+ \big[ a\varPhi \dot t_e - \left( r_e^2 + a^2 + a^2 \varPhi \right) \dot \phi_e \big] 
\dot \phi_e
\\
=&
E_e \dot t_e - L_e \dot \phi_e 
.
\end{split}
\end{equation}

The constants of motion can be calculated in terms of $r_e$ from equations 
\eqref{eq:rdot}, \eqref{eq:titadot} and \eqref{eq:ueue}; to do that we follow 
the discussion of \cite{Chandrasekhar:1983este} 
(pages 333-342 in Chapter 7, section 61).
Using \eqref{eq:titadot} in \eqref{eq:rdot} we note that the radial equation can 
also be expressed as:
\begin{equation}\label{eq:qubic+eq+r_e}
\begin{split}
0 =&
r_e^4 E_e^2 + r_e^2(a^2 E_e^2 - L_e^2 - \Delta_e ) + 2 M r_e (a E_e - L_e)^2  
,
\end{split}
\end{equation}
and defining 
\begin{equation}
u=1/r_e,
\end{equation}
and 
\begin{equation}\label{key}
Q_\pm = 1 - 3M u \pm 2 a \sqrt{M u^3}
;
\end{equation}
then, the values of the energy $E_e$ and angular momentum $L_e$ 
are given by\citep{Chandrasekhar:1983este}(p.335):
\begin{equation}\label{key}
E_e = 
\frac{1}{\sqrt{Q_\mp}}
\Big(
1 - 2 M u \mp a \sqrt{M u^3}
\Big) 
,
\end{equation}
and
\begin{equation}\label{key}
L_e =
\mp
\frac{\sqrt{M}}{\sqrt{u Q_\mp}}
\Big(
a^2 u^2 + 1 \pm 2 a\sqrt{M u^3}
\Big)
;
\end{equation}
where upper sing applies to retrograde orbits while lower sign applies to direct orbits.
After that, it only remains to compute the value of Carter's constant 
which is inferred from \eqref{eq:titadot}.

It should be noted that this procedure works until one finds roots of $Q$;
that for the case $M=1$, $a=0.98M$
is at approximately $u_0 = 0.806747578125$;
that is $r_0 = 1./0.806747578125 = 1.23954509082525$.
For a general value of $M$ but the same ratio for $a$;
one can see that $r_0$ scales with $M$.
Then we could use this procedure for all $r_e$ satisfying
\begin{equation}
M r_0 \leq r_e;
\end{equation}
and we  keep the values for $E_e$ and $L_e$
for the few cases $r_+ \le r < r_0 M$;
where it should be noted that
$r_+(M=1,a=0.98)=1.19899748742132$ and $r_+$ scales with $M$ as well.

Let us also note that for this unit value of the mass, $r_0$ coincides 
with $r_c$ which is defined as the unstable circular photon orbit on 
the equatorial plane. Let us recall that 
\citep{Chandrasekhar:1983este}
\begin{equation}\label{key}
r_c = 2 M \Bigg(
1 + \cos\left(
\frac{2}{3} \arccos\left(\pm \frac{a}{M}\right)
\right)
\Bigg)
;
\end{equation}
where upper sing applies to retrograde orbits while lower sign applies to direct orbits.
That is, the exact root for $Q$ is given by the analytical expression that gives
the position of the photon ring; which is going to be used as the critical value 
for the procedure presented above.
In our further computations, taking into account numerical safety, we 
will use this procedure for all $r \ge 1.01 r_c$, and we will keep the values for 
$E_c$ and $L_c$ for the few cases $r_+ \le r < 1.01 r_c$; since
a  particle in the accretion flow, will spend 
a very short time in this region, and will have a very small
contribution to the image we are calculating.
Our disk model will only deal with prograde motion since this choice is both natural for
a black hole in a stationary regime and numerical simulated images tends to disfavor the 
case of retrograde motions.

\subsection{The red/blueshift effects}
We finish the present section with the explicit calculation of the 
red/blueshift contribution due to the prograde circular motion of the
accretion disk model.
In order to do so,  
we must modulate our previous calculations of fluxes with the factor 
$1/(1+z)^4 = 1/(u_e^b \ell_b)^4$; where $u_e^a$ is the four-velocity of the emitter.
We have that
\begin{equation}\label{eq:ue_ell}
\begin{split}
g_{ab} u_e^a \ell^b =&
\left( 1 - \frac{2 M }{r_e}\right) \dot t_e \dot t  
+ \frac{2 a M}{r_e} (\dot t_e \dot\phi + \dot t \dot\phi_e)\\
&  
- \left( r_e^2 + a^2 + \frac{2 a^2 M}{r_e} \right) 
\dot\phi_e \dot\phi \\
=&
\bigg[
\left( 1 - \frac{2 M }{r_e}\right) \dot t_e  
+ \frac{2 a M}{r_e}  \dot\phi_e
\bigg] \dot t \\
&-
\bigg[
-\frac{2 a M}{r_e} \dot t_e  
+ \left( r_e^2 + a^2 + \frac{2 a^2 M}{r_e} \right) 
\dot\phi_e
\bigg] \dot{\phi} \\
=&
E_e \dot t - L_e \dot \phi 
.
\end{split}
\end{equation}
Here, for numerical convenience we choose to use the conservation equation 
of the circular orbits.
So that the factor that takes into account the gravitational
red/blueshift effects is simply $1/(E_e \dot t - L_e \dot \phi)^4$,
evaluated at the position of the emission,
and where $\dot t$ and $\dot \phi$ corresponds to the
coordinate velocities of the photon.

The result of these calculation can be seen in the graphs of section \ref{sec:Results}.

\section{Numerical implementation}\label{sec:Numerical+implementaion}

\subsection{Description of the code}

In this work we use a numerical implementation that performs the integration of 
the combined systems given by equations \eqref{eq:ell+nabla+ell} and \eqref{eq:geod+dev+equation} 
on the Kerr spacetime.
It was previously mentioned in the introduction that our method provides an enhanced 
prescription in comparison with the previous pioneering work of references 
\cite{Pineault1977ApJ, Pineault1977ApJ_II} and recent modifications of it \citep{James:2015yla}.
More precisely, in those previous works the Weyl curvature scalar $\Psi_{0}$ was expressed as the 
product of a real function $\chi$ and another complex one $\Psi_{0*}$ in the form 
$\Psi_{0} = e^{-i2\chi} \Psi_{0*}$; where $\chi$ depends on the propagation properties of the 
complex null vector $m^a$ and have to be found through integration of equation (14) of 
reference \cite{Pineault1977ApJ}, while $\Psi_{0*}$ is given in terms of at least 14 auxiliary
functions of the coordinates and the velocities (see equations (15), (16), (17) and appendix 
in the same reference). 
One can see that the presentation of the formulae needed to compute $\Psi_{0*}$
requires two pages in \cite{James:2015yla}, which indicates 
a proportional amount of computation
for each evaluation of this Weyl curvature component
In contrast, our expression for $\Psi_{0}$, namely equation \eqref{eq:Psi0} reduces 
to a single simple 
function of two coordinates along the null geodesics.
Since $\Psi_{0}$ is used about $n$ times at each step of integration,
of the null geodesic deviation equations, to obtain a $n$th order precision,
one concludes that our expression represents a huge improvement
in the efficiency of the computation of the optical scalars
in this problem.

The set of equations in its suitable form for numerical integration and the appropriated
initial conditions are presented below in subsection \ref{ap:Linear+order+system}.
The initial conditions were set in a polar grid associated with the lines of sight within 
a small portion of the sky at the position of the observer. 
Typical calculations involve 700 target points, used to generate
an square image  of $120\mu$as in size.

The code works by running the integration from the observer position until one of the 
following conditions takes place; the null geodesic: $i)$ crosses the event horizon, 
$ii)$ hits the equatorial plane in the regions where the accretion disk is defined or, 
$iii)$ hits the equatorial plane in the outer regions.

Let us note that the above procedure implies that we are
 considering 
that the null geodesics are emanating from the equatorial
accretion disk; which does not exclude the
possible presence of a thick and more tenuous disk which is 
optically thin surrounding the BH.

We wrote two independent codes in {\sc Fortran90}, one was run in double
precision and the other in quadruple precision, making use of a 7-8 Runge-Kutta solver from 
the suite RKSuite\citep{rksuite-90} both of them producing similar results.

The relative errors $\epsilon_{C}$ of several available conserved quantities
have been employed as a method for error control of the integration 
procedure.
The typical values for a broad range of parameters are of order
$\epsilon_{C} \sim 10^{-20}$, in the quadruple precision version; 
which gives great reliability of the code.

\subsection{Linear order system}\label{ap:Linear+order+system}
In order to integrate equations \eqref{eq:ell+nabla+ell} 
and \eqref{eq:geod+dev+equation};
one can take  different approaches to cast the system of equations 
in a first order system well suitable for numerical integration.
The different approaches vary in the way one handle the geodesic equations; 
namely by using its second order version such as are given by \eqref{eq:ell+nabla+ell},
using the first order version presented in \ref{subsec:Null+geod+Kerr} or a combination
of both.
Below is our choice for this work: 
\begin{align}
\dot t 
=& 
\frac{1}{\Sigma \Delta}\left[ E \Big(
(r^2 + a^2)^2 - \Delta \, a^2 \sin(\theta)^2 \Big)
- 2 a M r L \right]
, 
\\
\dot{r} =& v^r,
\\
\dot{v}^r =&  
(v^{\theta})^2 r \frac{\Delta}{\Sigma} 
+ \frac{a^2}{\Sigma} \dot{r} \dot{\theta} \sin(2\theta) 
- (v^{r})^2 \frac{r \Delta + \left(M - r \right)\Sigma}{\Sigma \Delta} \nonumber 
\\
& 
- \dot{t}^2 
\frac{ M \Delta \big( r^2 - a^2 \cos(\theta)^2 \big)}{\Sigma^3} \nonumber
\\
& + 2 \dot{t}\dot{\phi}
\frac{ a M \Delta \big( r^2 - a^2 \cos(\theta)^2 \big) \sin(\theta)^2}{\Sigma^3} \nonumber
\\
&
+
\dot{\phi}^2 
\frac{\Delta \sin(\theta)^2}{\Sigma^3}
\Big( r\Sigma^2 \\
& \qquad \qquad \qquad \;\; - a^2 M\big( r^2 - a^2\cos(\theta)^2\big)\sin(\theta)^2 \Big)
,
\\
\dot{\theta} =& v^\theta, 
\\
\dot{v}^\theta =& \frac{a^2 M r \sin(2\theta)}{\Sigma^3} \dot{t}^2 
- \frac{2aMr \left( r^2 + a^2 \right)\sin(2\theta)}{\Sigma^3}\dot{t}\dot{\phi} \nonumber 
\\
& - \frac{a^2 \sin(2\theta)}{2 \Sigma \Delta} (v^{r})^2 
- \frac{2r}{\Sigma}\dot{r}\dot{\theta} +  
\frac{a^2 \sin(2\theta)}{2 \Sigma} (v^{\theta})^2  \nonumber
\\
&+ \frac{\sin(2\theta)}{2\Sigma^3} 
\Big( \left( r^2 + a^2\right)\Sigma^2 
\\
& \qquad \qquad \quad + 2a^2rM\sin(\theta)^2 
\left( r^2 + a^2 + \Sigma \right) \Big) \dot{\phi}^2
,
\\
\dot \phi =& \frac{1}{\Sigma\Delta}\left[ 2E a M r + (\Sigma - 2Mr) \frac{L}{\sin(\theta)^2}\right]
,
\end{align}
\begin{align}
\dot{\varsigma}_{R1}^{} =& \upsilon^{\varsigma}_{R1}, 
\\
\dot{\upsilon}^{\varsigma}_{R1} =& - \varsigma_{R1}^{} \Psi_{0R}^{} - \varsigma_{I1}^{} \Psi_{0I}^{},
\\
\dot{\varsigma}_{I1}^{} =& \upsilon^{\varsigma}_{I1},
\\
\dot{\upsilon}^{\varsigma}_{I1} =& - \varsigma_{R1}^{} \Psi_{0I}^{} + \varsigma_{I1}^{} \Psi_{0R}^{},
\end{align}
\begin{align}
\dot{\varsigma}_{R2}^{} =& \upsilon^{\varsigma}_{R2}, 
\\
\dot{\upsilon}^{\varsigma}_{R2} =& - \varsigma_{R2}^{} \Psi_{0R}^{} - \varsigma_{I2}^{} \Psi_{0I}^{},
\\
\dot{\varsigma}_{I2}^{} =& \upsilon^{\varsigma}_{I2},
\\
\dot{\upsilon}^{\varsigma}_{I2} =& - \varsigma_{R2}^{} \Psi_{0I}^{} + \varsigma_{I2}^{} \Psi_{0R}^{};
\end{align}
with the following initial conditions:
\begin{align}
t_0 =& t_o = 0, \\
r_0 =& r_o, \\
v^r_0 =& - \frac{\sqrt{\mathcal{R}(r_o)}}{\Sigma_o} , \\
\theta_0 =& \theta_o, \\
\ell^\theta_0 =& \pm \frac{\sqrt{\varTheta(\theta_o)}}{\Sigma_o}, \\
\phi_0 =& \phi_o = - \frac{\pi}{2}
,
\end{align}
\begin{align}
\varsigma_{R1_0} =& 0 , \\
\upsilon^\varsigma_{R1_0} =& 1, \\
\varsigma_{I1_0} =& 0 , \\
\upsilon^\varsigma_{I1_0} =& 0
,
\end{align}
\begin{align}
\varsigma_{R2_0} =& 0 , \\
\upsilon^\varsigma_{R2_0} =& 0, \\
\varsigma_{I2_0} =& 0 , \\
\upsilon^\varsigma_{I2_0} =& 1
.
\end{align}

\subsection{Blurring}
Besides of the optical effects due to the background geometry and 
the emission model of the accretion flow 
in the surroundings of the SMBH, 
simulated images also have to take into consideration the limitations introduced by the instrumental 
array and any other possible noise in the way towards the final 
image reconstruction.

For radio observations of very long base interferometry such as those 
needed for the study of the nearest region of SMBH's, the main contribution in 
absence of instrumental noise is due to the intrinsic resolution of the array 
of antennas. 
For the EHT the maximum theoretical resolution of the array during its observational 
campaign on April 2017 at wavelengths of $\sim 1.3\text{mm}$\citep{Akiyama:2019cqa, Akiyama:2019brx}) 
was $\sim 25 \mu\text{as}$. 
We will include this effect by blurring the images with a circular Gaussian filter 
characterized by a full width at half maximum (FWHM) with values closer to the 
employed by the EHT in its three internal pipelines \texttt{eht-imaging, SMILI}, and \texttt{DIFMAP}.
The values employed for the FWHM were respectively, 17.1, 18.6 and 20 according to 
\cite{Akiyama:2019bqs}.
Let us recall that $\text{FWHM} = 2\sqrt{2\ln 2} \,\sigma$; with $\sigma$ the mean square root of 
the Gaussian function.
Our graphs use the value $dx=11$ for the typical distance $dx=\sqrt{2} \sigma$ used in the Gaussian 
blurring in the {\sc gnuplot} graphic tools; that corresponds to a FWHM$=18.3$.

\section{
Simulated images with gravitational lens magnification and red/blueshift effects
}\label{sec:Results}

\subsection{Basic model}

This section contains our final images that were 
simulated with the numerical approach presented in this article and 
the previous setting described in section \ref{sec:accretion+disk}.

Below, we list the parameters that define our base model for the black hole,
the distance to M87 and the two temperature thin disk model.
For the mass of the supermassive black hole  we take the value,
estimated in \cite{Gebhardt:2011yw}, of $(6.6\pm0.4) \times 10^{9}M_\odot$.
For the intrinsic angular momentum of the black hole
we take the value, estimated in \cite{Feng:2017vba}, of 
$0.98^{+0.012}_{-0.02}$ 
of the total mass.
We corroborate in this work that this value provides 
an excellent parameter for our simple model.
We assumed a distance from the observer to the SMBH equal to those of the 
usual values reported for its hosting galaxy M87, namely $d_l=16.7$Mpc.

Our base model is a two temperature thin disk inspired in a model as depicted 
in the central graph of fig. 2 of reference \cite{Moscibrodzka:2015pda};
taking two representative values for the temperatures in the equatorial 
plane that are compatible with those employed in that reference.
In dimensionless units, in terms of $\Theta_e = k_B T / m_e c^2$,
the values employed were:
an inner temperature disk of $\Theta_e=7$, 
an external temperature disk of $\Theta_e=0.2$,
an ambient temperature of $\Theta_e=0.05$
and a black hole temperature set to $\Theta_e=0$.

The simple configuration we are considering is:
i) an inner disk that expands in the range $r_+< r < r_\text{in}$,
   where $r_+$ is the radius of the event horizon \eqref{eq:r+horizon};
ii) an intermediate disk in the range $r_\text{in} < r <r_\text{ex}$;
iii) an external disk with $r_\text{ext} < r$.
The inner radius $r_\text{in}$ is equivalent to 
a 16$\mu$as extension for the $d_l$ distance mentioned above;
while the external radius $r_\text{ex}$ is equivalent to 
a 45$\mu$as angle in the sky of the observer.

The spin of the black hole, its inclination with respect to the line of sight and
the position angle (PA) of its projection will be
indicated in each of the cases presented next.
It is probably worthwhile to realize that the most salient features associated with M87, 
namely its large scale jet, constitutes a guide a priori that suggests the most probable 
orientation of the spin of the BH and its PA, and therefore we 
will only consider the aligned and anti-aligned configurations between the jet
and the spin.
For the large scale observed jet, we have used as PA the value that according to 
\cite{Kovalev:2007hy}, is 290 degrees (east of north).
This is almost in agreement with EHT assumption\citep{Akiyama:2019fyp}, where they used
the value of 288 degrees.

Regarding the color palette, we have used a variation of the AFM hot
{\sc gnuplot} palette, adapted to obtain a similar pattern
as shown in the EHT images.

\subsection{Graphs of the two temperature disk model}\label{subsec:2T+model}

\begin{figure*}
\includegraphics[clip,width=0.495\textwidth]{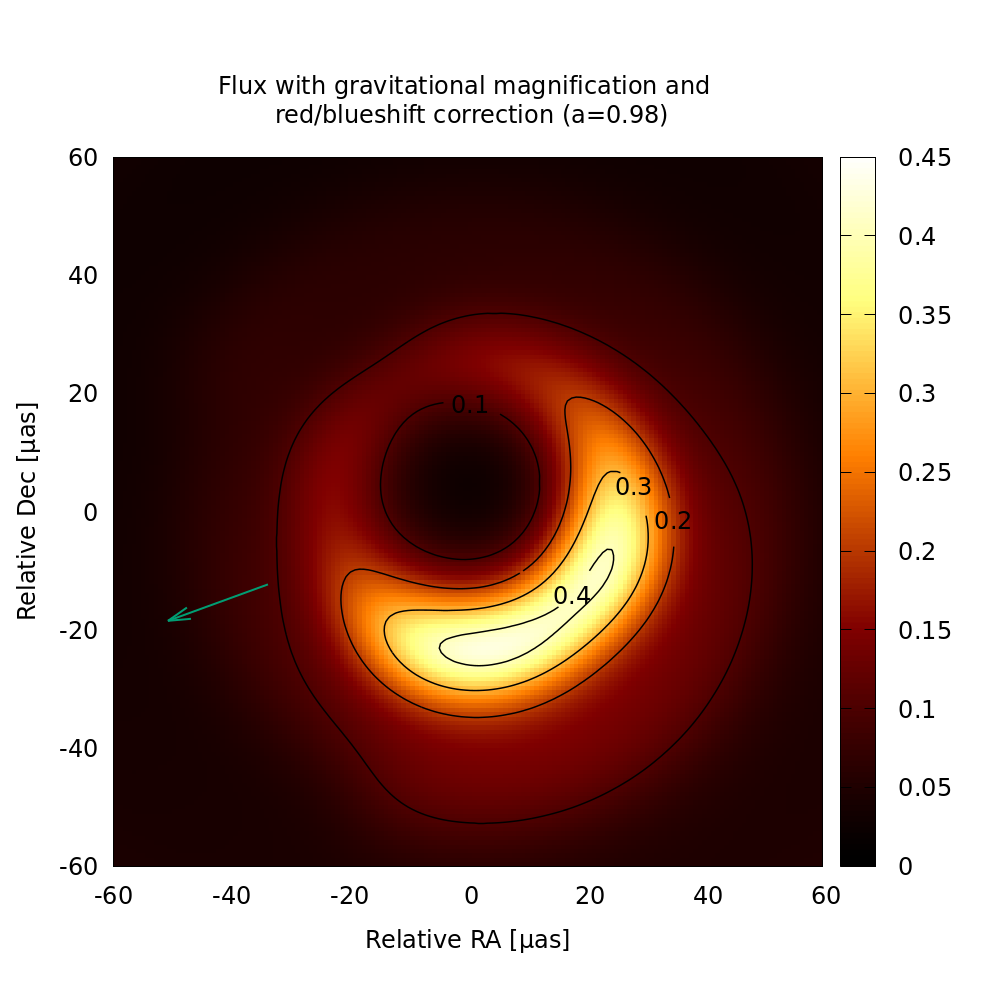}
\includegraphics[clip,width=0.495\textwidth]{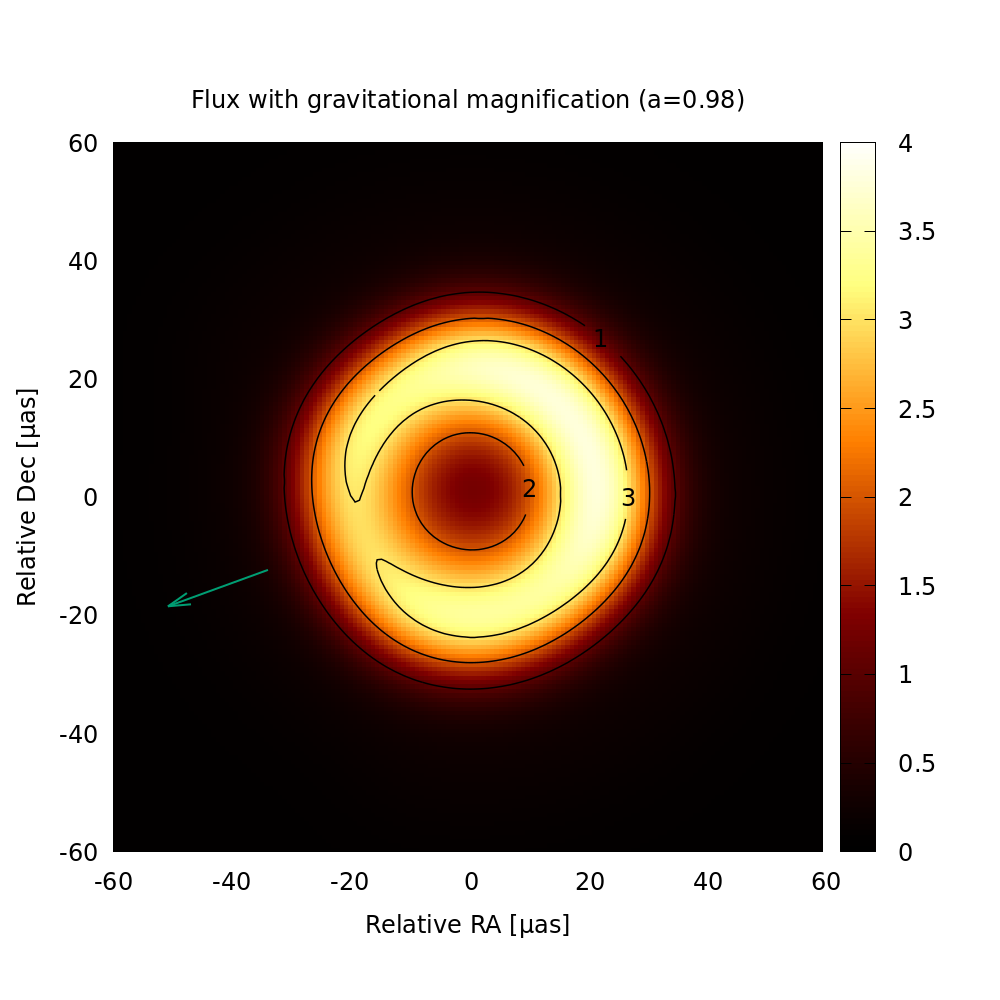}
\caption{ Anti-aligned configuration:
Left graph shows the flux for a two temperature disk
model, modulated by the gravitational lens magnification, 
the red/blueshift correction (due to gravity and motion)
and smoothed by a Gaussian function.
Right graph shows the flux for the same
model, but only modulated by the gravitational lens magnification
and smoothed by a Gaussian function.
The angular momentum $(a=0.98M)$ of the black hole and the disk are opposite to the jet.
Here we use the astrophysical angular coordinates.	
The arrow shows the projected direction of the angular momentum.
}
\label{fig:flux-arrow-2}
\end{figure*}

We present here our results for the simplest case of the two temperature thin disk 
model as described previously.
In Fig. \ref{fig:flux-arrow-2} it is shown the flux that 
one would observe for this model
when both the angular momentum of the SMBH and the accretion disk are opposite to 
the observed jet.
Please note that we use $a>0$ values to indicate angular momentums that
are anti-aligned with the observed jet.

\begin{figure*}
\centering
\includegraphics[clip,width=0.495\textwidth]{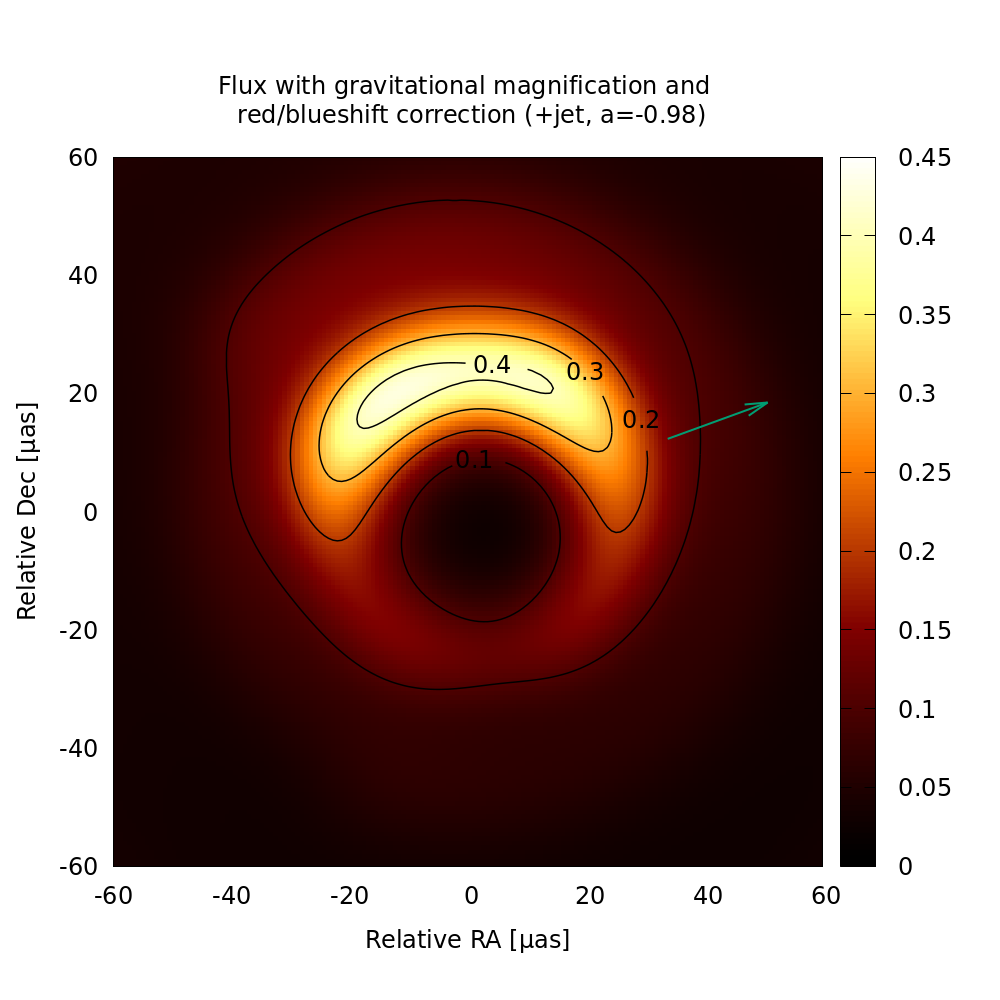}
\includegraphics[clip,width=0.495\textwidth]{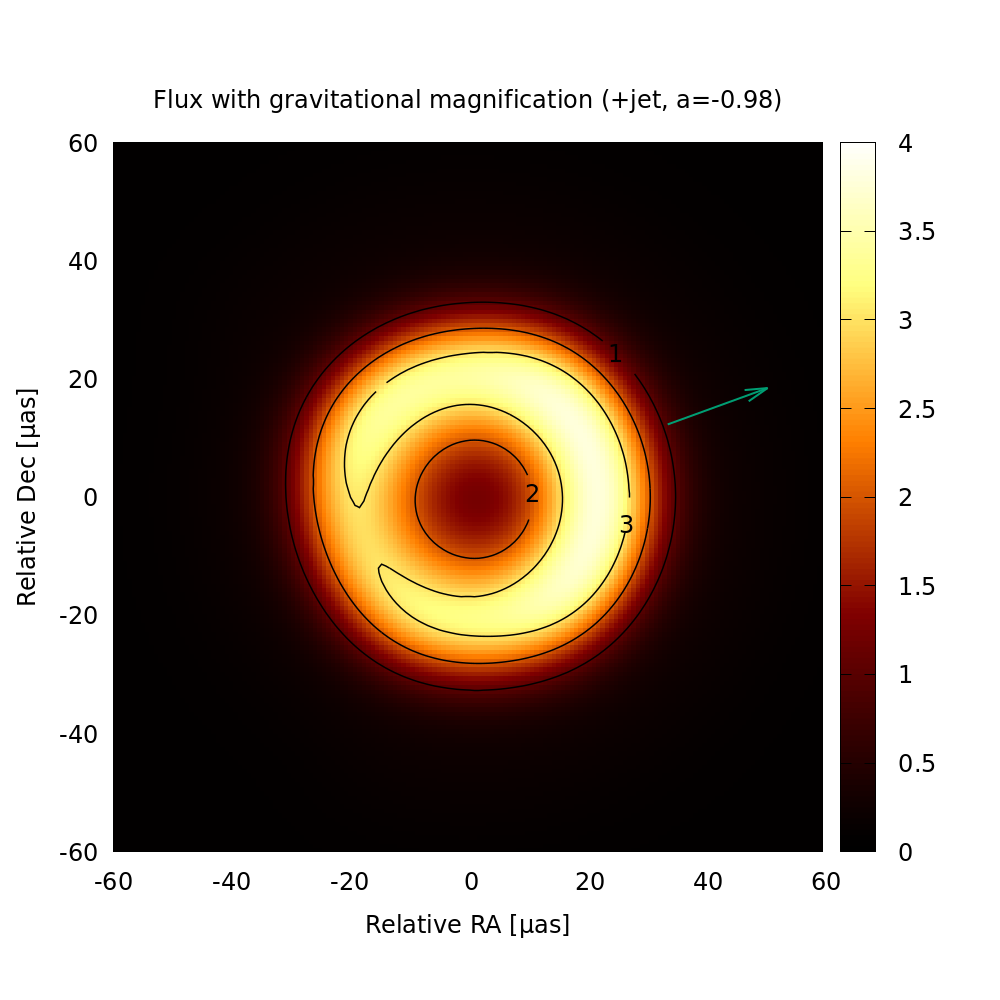}
\caption{Aligned configuration: Left graph shows the flux for a two temperature disk
model, modulated by the gravitational lens magnification,
the red/blueshift correction (due to gravity and motion)
and smoothed by a Gaussian function.
Right graph shows the flux for the same
model, but only modulated by the gravitational lens magnification
and smoothed by a Gaussian function.
The angular momentum $(a=-0.98 M)$ of the black hole and the disk are in the same direction as the jet.
Here we use the astrophysical angular coordinates.	
The arrow shows the projected direction of the angular momentum; 
in this case the PA of jet 
is the same as that of the spin of the black hole.
}
\label{fig:flux-arrow-4}
\end{figure*}

Fig. \ref{fig:flux-arrow-4} corresponds to the angular 
momentum of the SMBH and the accretion disk aligned with the jet.
The left panels in both figures contains the red/blueshift effect due to the 
geodesic motion of the plasma together with the contribution coming from the 
magnification. To illustrate this last contribution, in the right panels we have built
images only modulated by the magnification factor \eqref{eq:magnification}.
One can notice from that figures that the 
dominant effect is due to the red/blueshift modulation.

One can observe up to this point,
considering these two choices,
that the left graph in Fig. \ref{fig:flux-arrow-2}
produces an enhanced brightness of a crescent
shape similar to those in the south-west location part of the reported images by the EHT 
Collaboration (see for example Fig. \ref{fig:EHT+April_11} again, or 
	Fig. \ref{fig:las+tres} and Fig. \ref{fig:final} below).
This supports the choice of the anti-aligned orientation for the angular momentum
of the SMBH in the center of M87.

For the image of April 11 reproduced in Fig. \ref{fig:EHT+April_11} the bright crescent sector 
at the south, has its mean at a PA $\simeq 170^o$ (see section 5 of \cite{Akiyama:2019cqa} and table 7 of 
\cite{Akiyama:2019bqs}) with an approximate extension between PA $\in (80^o,270^o)$ 
(Fig. \ref{fig:EHT+April_11} and fig. 26 and 27 of \cite{Akiyama:2019bqs}).
The diameter has been estimated as $d \simeq 41\mu$as 
with a width $w \lesssim 20\mu$as (section 9.3 in \cite{Akiyama:2019bqs}). 
Our images presented in this section, in particular that of the left panel of Fig. \ref{fig:flux-arrow-2}
shares a similar estimated values for the width and diameter of the asymmetric bright
ring structure. Instead, the mean position of this structure is located at PA $\sim 220^o$ with 
an approximate extension between PA $\in (135^o,290^o)$.
Then, the extension in this case is slightly smaller than in Fig. \ref{fig:EHT+April_11}
and the value of the PA for the mean point has an appreciable change.
It is important then, to recall that we have fixed the orientation of the spin of the SMBH
in the opposite direction of the jet and so, we are not allowing to relax the observed tension 
by this disagreement just by changing the orientation of the spin of the BH.
It is for this reason that we explore modifications to this model in
forthcoming subsections.
Nevertheless, it is interesting to note that this tension significantly improves if one
would consider the values for the PA indicated for the black hole spin in section 5 of 
reference \cite{Akiyama:2019fyp} where an estimation of the parameters for the black
hole is done over a large image library employed by the Collaboration.

\subsection{The effect of the spin parameter on the two temperature disk model}\label{subsec:2T+spinvarying}

\begin{figure*}
\centering
\includegraphics[clip,width=0.495\textwidth]{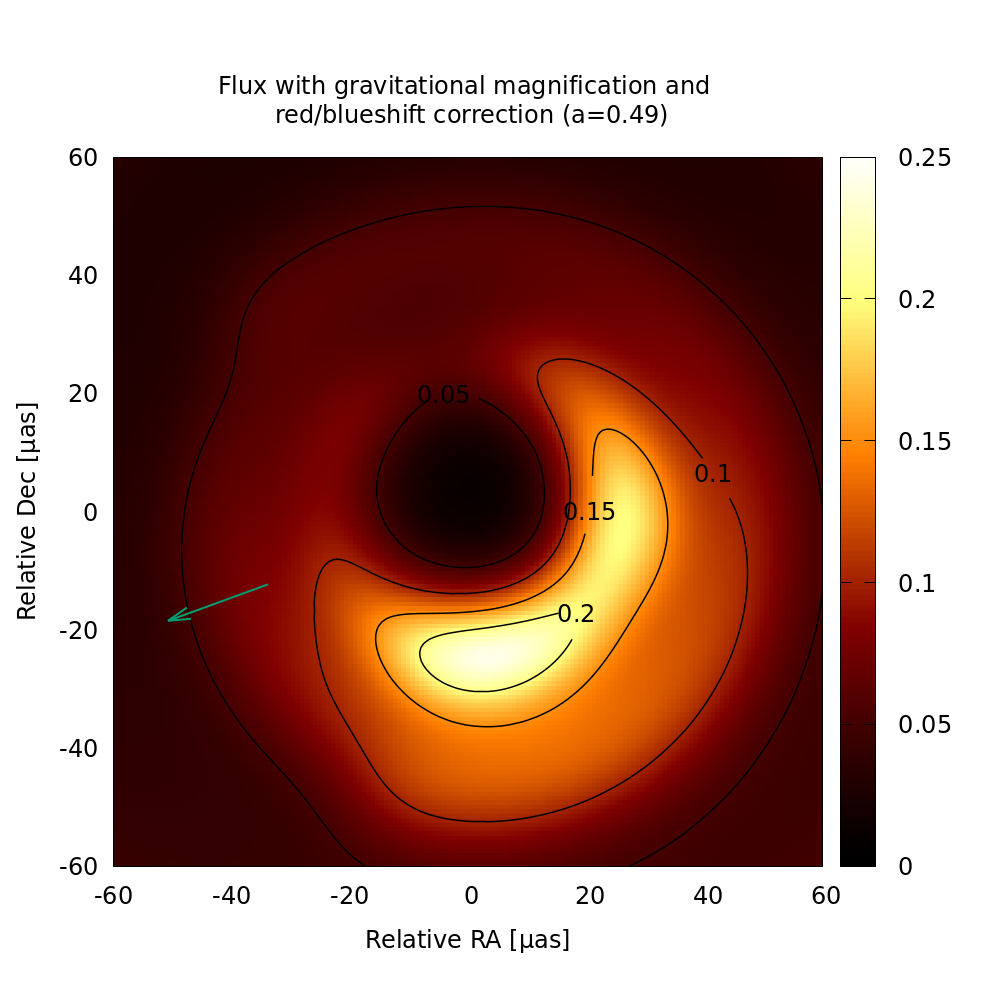}
\includegraphics[clip,width=0.495\textwidth]{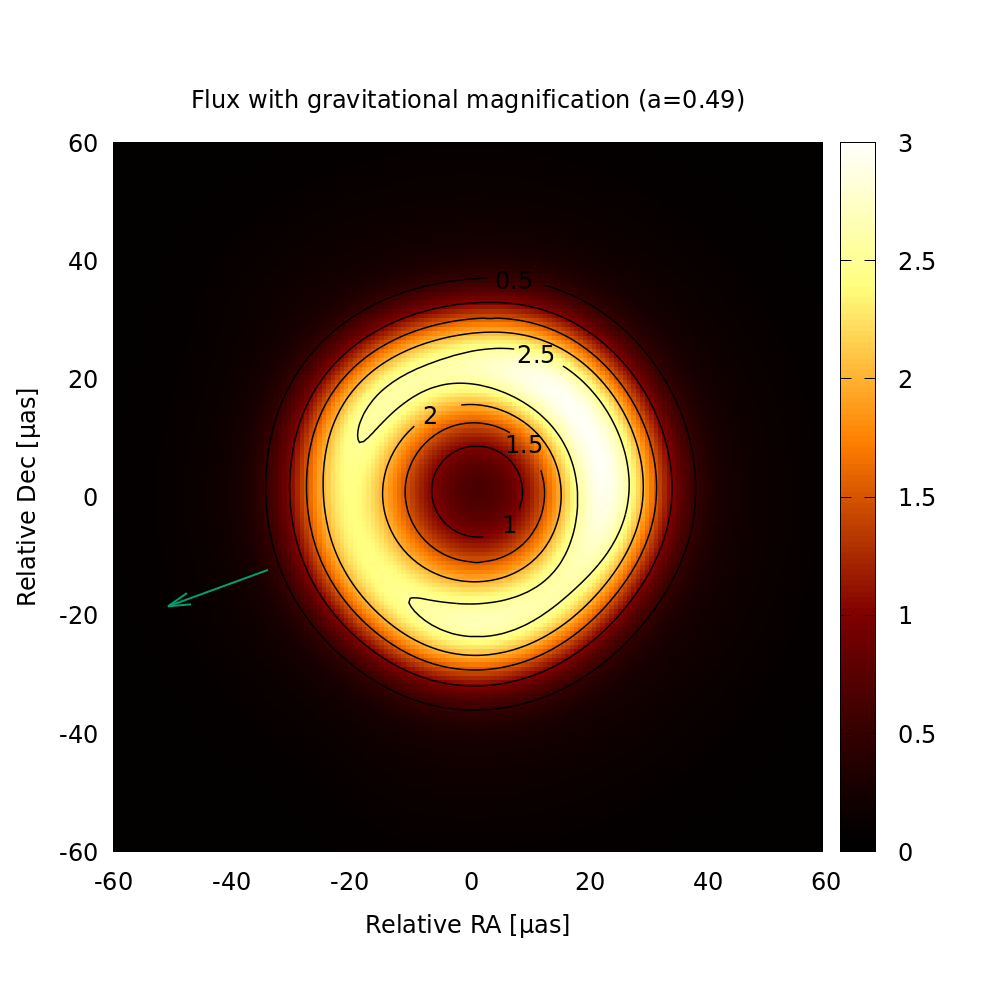}
\caption{Intermediate spin values: Left graph shows the flux for a two temperature disk
model, modulated by the gravitational lens magnification, 
the red/blueshift correction (due to gravity and motion)
and smoothed by a Gaussian function.
Right graph shows the flux for a two temperature disk
model, modulated by the gravitational lens magnification
and smoothed by a Gaussian function.
The angular momentum $(a=0.49M)$ of the black and the disk are opposite to the jet.
Here we use the astrophysical angular coordinates.	
The arrow shows the projected direction of the angular momentum.
}
\label{fig:flux-arrow-21}
\end{figure*}

\begin{figure*}
\centering
\includegraphics[clip,width=0.495\textwidth]{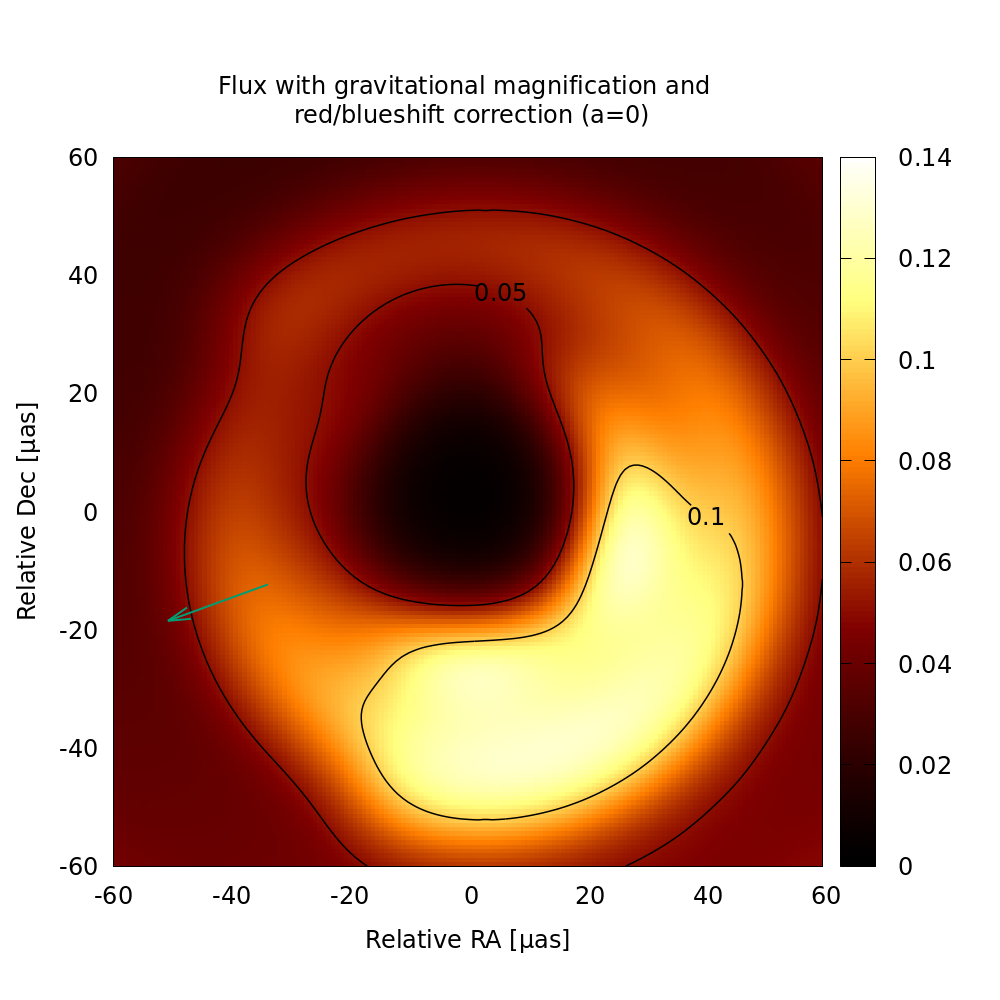}
\includegraphics[clip,width=0.495\textwidth]{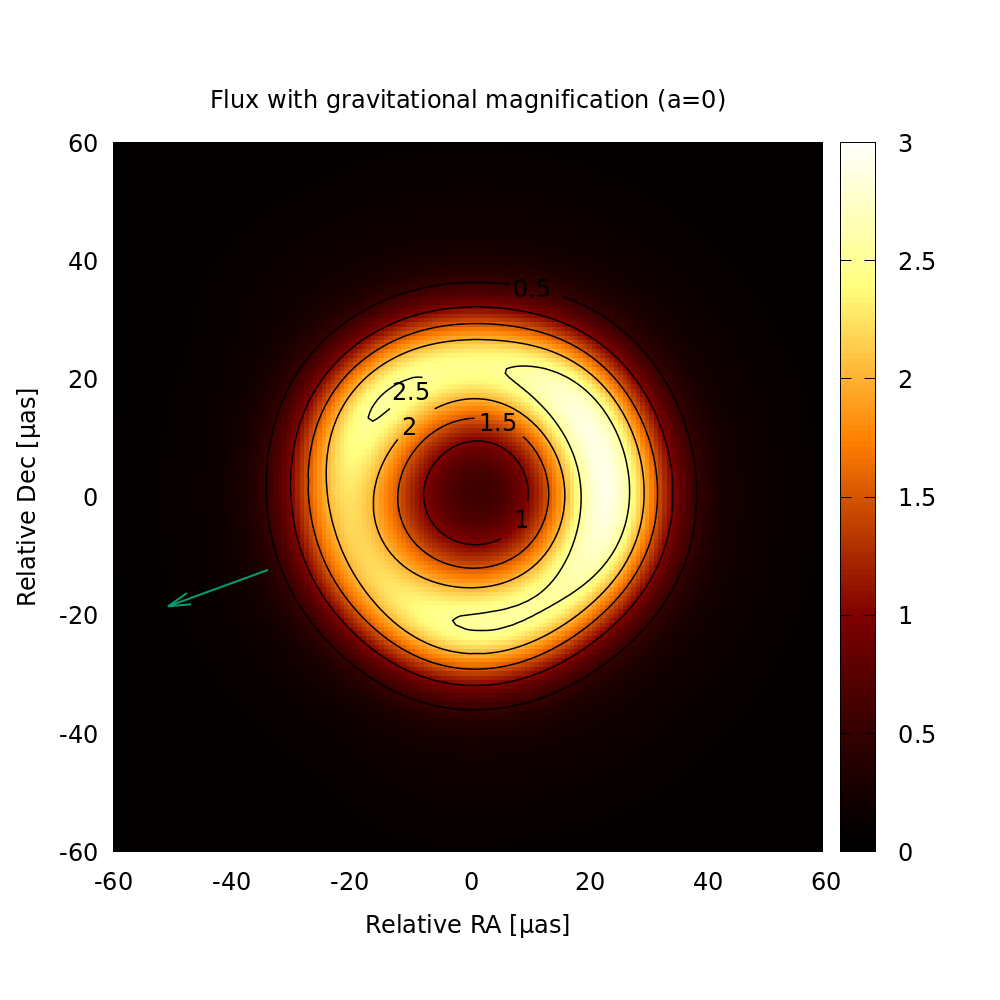}
\caption{Zero angular momentum: Left graph shows the flux for a two temperature disk
model, modulated by the gravitational lens magnification, 
the red/blueshift correction (due to gravity and motion)
and smoothed by a Gaussian function.
Right graph shows the flux for a two temperature disk
model, modulated by the gravitational lens magnification
and smoothed by a Gaussian function.
The angular momentum of the black hole is zero and 
the angular momentum of the disk is opposite to the jet.
Here we use the astrophysical angular coordinates.	
The arrow shows the projected direction of the angular momentum.
}
\label{fig:flux-arrow-6}
\end{figure*}

Before embarking in a modification of the previous model it is interesting to 
explore further changes on the images produced by variations 
of magnitude of the angular momentum of the 
geometry. 
So, we here consider changes associated with a two 
temperature thin disk model when we vary the magnitude of the angular momentum of the SMBH and we study
its impact on the images through the modification of the kinematics redshift and the magnification 
effects.
For that purposes we study, in addition to the cases presented before,
two situations: i) the extreme case of a SMBH with 
zero angular momentum and, with an 
accretion disk with its angular momentum anti-aligned with respect to the jet, shown in 
Fig. \ref{fig:flux-arrow-6} and ii) the intermediate case of a SMBH with  $a/M = 0.49$ 
and pointing outward to the jet, shown in Fig. \ref{fig:flux-arrow-21}.

The main features that appear due to the decrease in angular momentum 
include a decrease in the observed intensity; again
this effect is mainly dominated by the variation 
produced by the red/blueshift effect on 
the accretion disk.
It can also be observed a widening of the brightest zone.
Therefore, from Figs. \ref{fig:flux-arrow-21} and \ref{fig:flux-arrow-6}
we conclude that low values of angular momentum are
disfavored with respect to the most probable ones indicated in the previous subsection. 

\subsection{Two temperature disk model with a bar}

\begin{figure*}
\centering
\includegraphics[clip,width=0.495\textwidth]{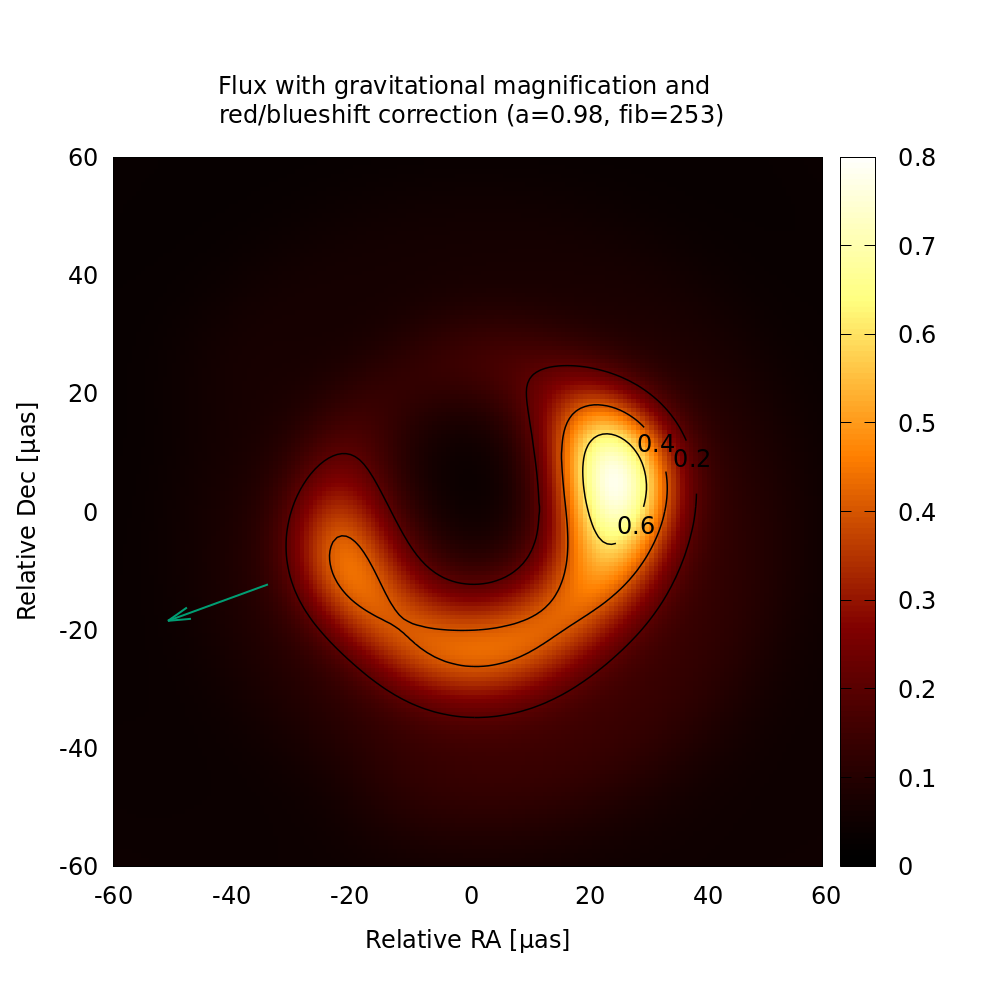}
\includegraphics[clip,width=0.495\textwidth]{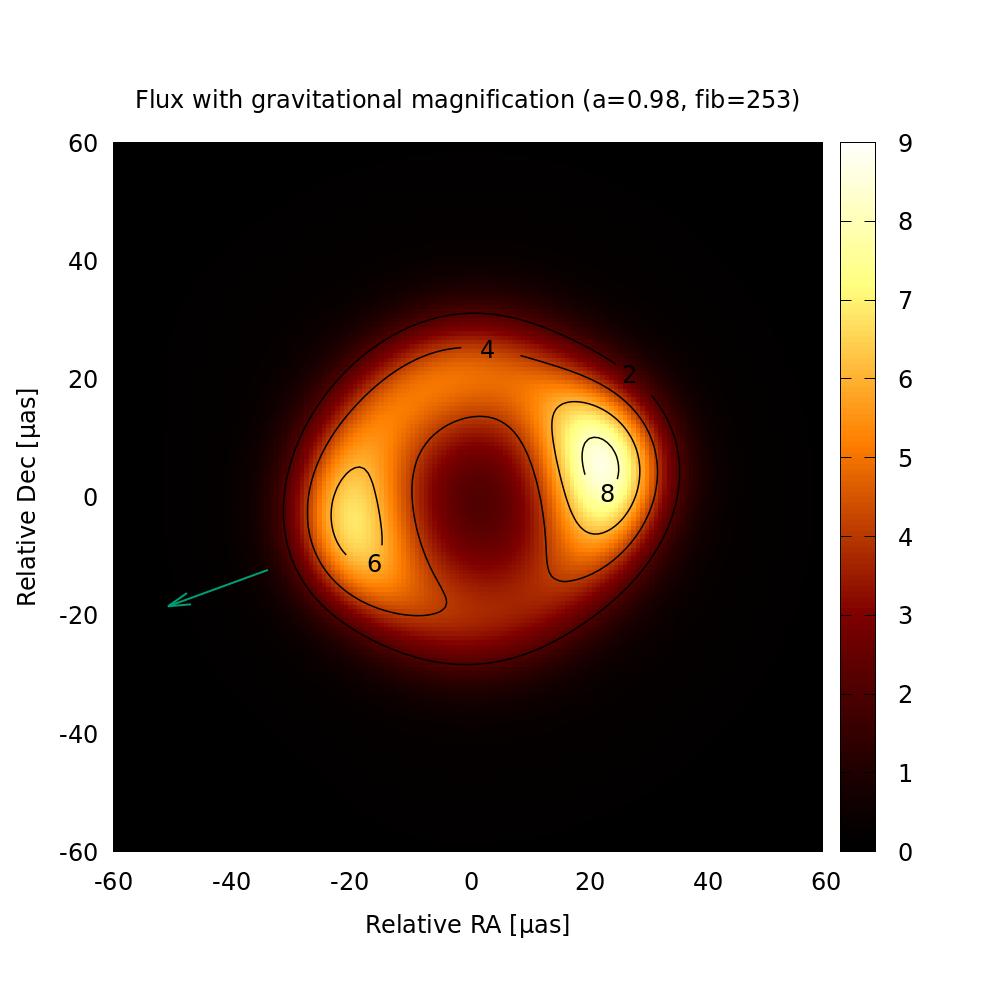}
\caption{Left graph shows the flux for a two temperature disk
model and a symmetric two sides bar, modulated by the gravitational lens magnification,
the red/blueshift correction (due to gravity and motion)
and smoothed by a Gaussian function.
Right graph shows the flux for a two temperature disk
model and a symmetric two sides bar, modulated by the gravitational lens magnification
and smoothed by a Gaussian function.
The angular momentum (a=0.98) of the black hole and the disk are in the opposite direction of the jet.
Here we use the astrophysical angular coordinates.	
The arrow shows the projected direction of the angular momentum.
}
\label{fig:flux-arrow-61}
\end{figure*}

\begin{figure*}
\centering
\includegraphics[clip,width=0.495\textwidth]{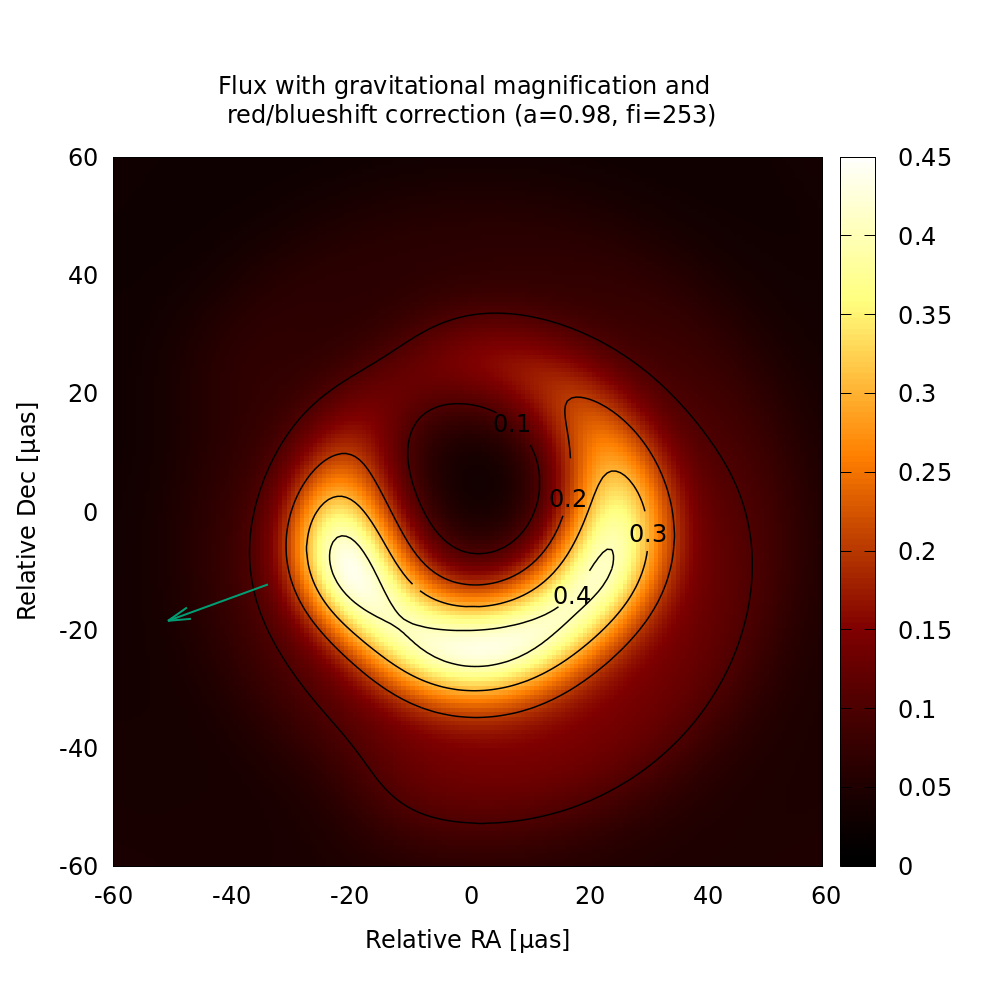}
\includegraphics[clip,width=0.495\textwidth]{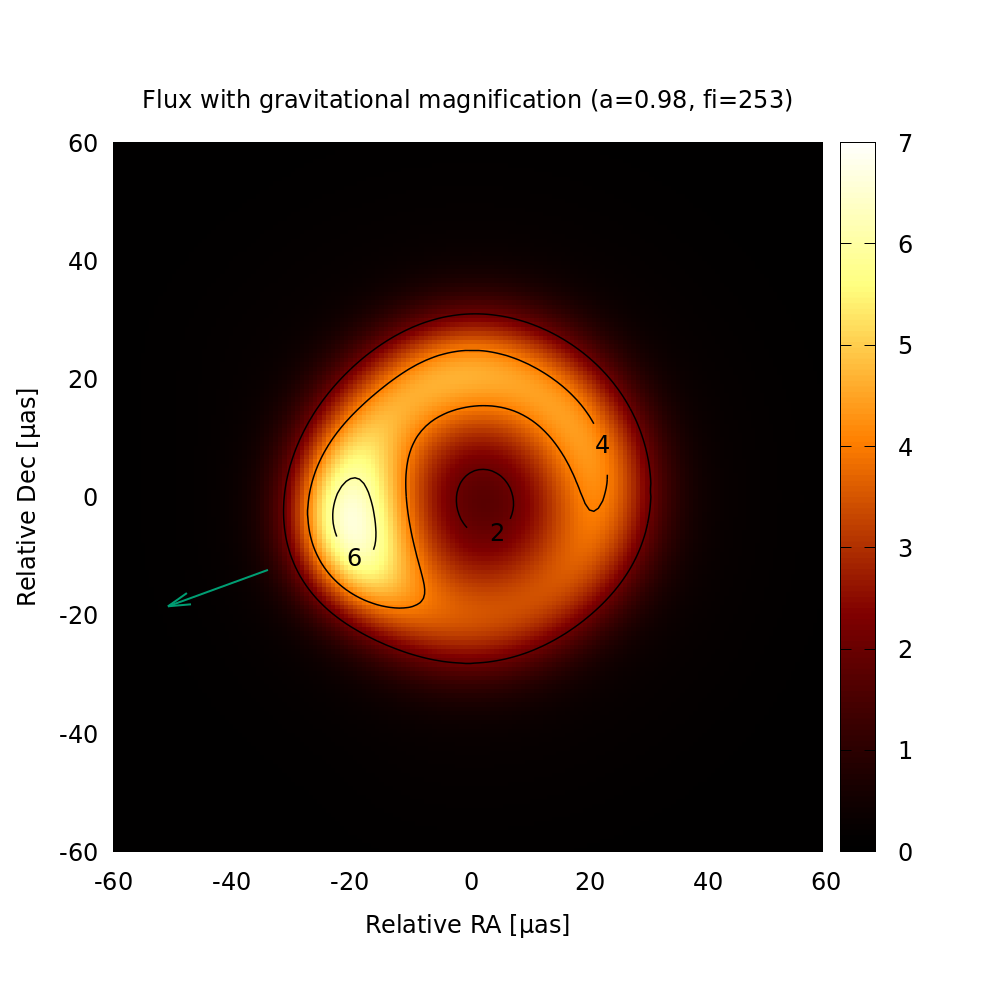}
\caption{Left graph shows the flux for a two temperature disk
model and a one side bar, modulated by the gravitational lens magnification,
the red/blueshift correction (due to gravity and motion)
and smoothed by a Gaussian function.
Right graph shows the flux for a two temperature disk
model and a one side bar, modulated by the gravitational lens magnification
and smoothed by a Gaussian function.
The angular momentum (a=0.98) of the black hole and the disk are in the opposite direction of the jet.
Here we use the astrophysical angular coordinates.	
The arrow shows the projected direction of the angular momentum.
}
\label{fig:flux-arrow-5}
\end{figure*}

In subsection \ref{subsec:2T+model} we have shown that the
two temperature thin disk model produces images
with similar features to those of the EHT with a SMBH near to a extreme Kerr.
In particular, it seems to give account of the south-west crescent with a width $w$ and diameter
parameters consistent with the images of April 11.
However, in Fig. \ref{fig:EHT+April_11} one can also observe two bright peaks in the 
crescent, one at the south-west and a smaller one at east position.
To give account of both features, we included in our model an emission component 
intended to balance the lack of brightness at the east location.
Perhaps the most simple modification to achieve this goal 
is to add a bar like structure at 
appropriate angular direction at a higher temperature.
We have chosen the bar to have the temperature $\Theta_e=56$, with a representative
width angle of 15$^\circ$.
The orientation angle and width of the bar were adjusted to show similarity with the 
observational images. 
A symmetric (two side) bar that includes emissions at angles $\phi = 73^o$
and $\phi = 253^o$, is presented in Fig. \ref{fig:flux-arrow-61}.
Since the symmetric, two side bar does not generate the desired image, 
we have also
considered an asymmetric, one side bar, at the reference angle $\phi = 253^o$,
whose image is shown in Fig. \ref{fig:flux-arrow-5}.

From this figure one can observe that the asymmetric configuration produces a quite 
impressive result with the desired features that were absent in the previous images, namely 
an extended bright crescent from south-west to south-east with level sets resembling the 
presence of two well distinguished maxima connected between them.
On the contrary the symmetric bar configuration gives origin to an undesired patter of flux
that draws away from the observational images.

\begin{figure*}
\centering
\includegraphics[clip,width=0.95\textwidth]{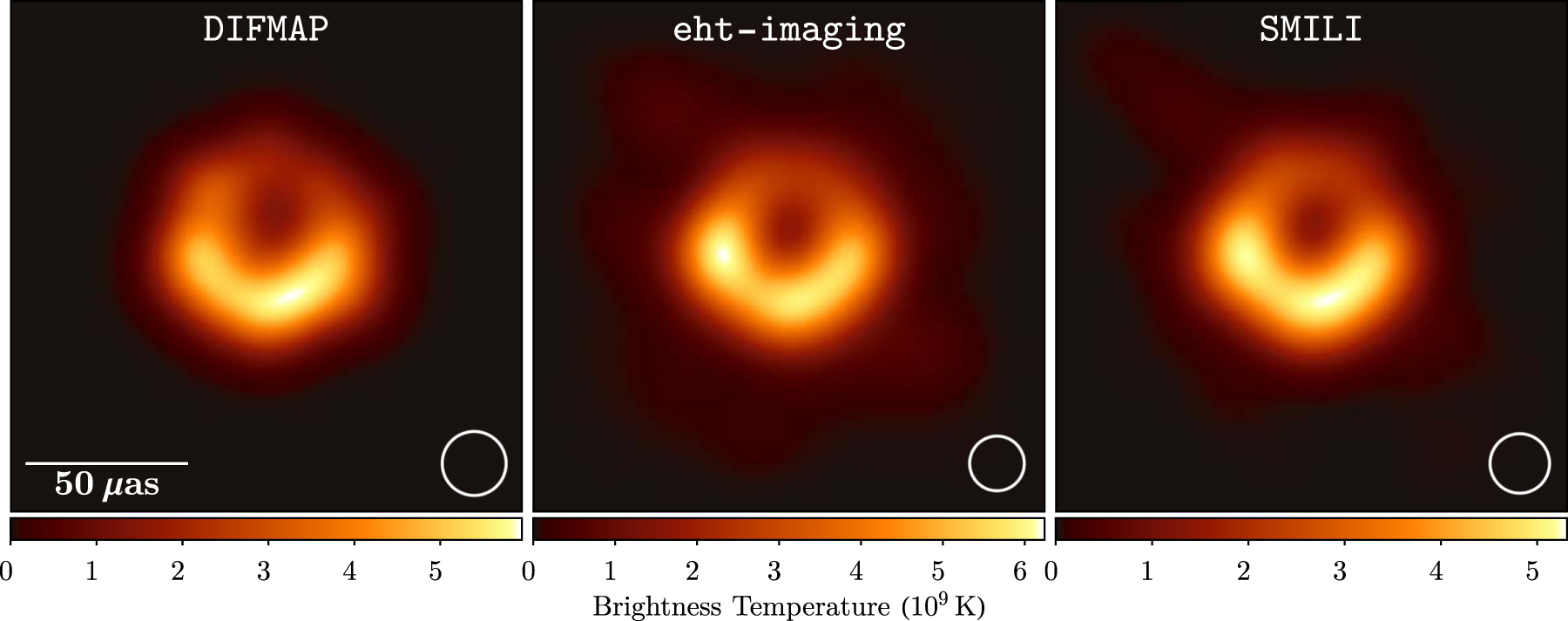}
\caption{
	We reproduce here the three EHT fiducial images of the innermost part of M87
constructed with the pipelines \texttt{DIFMAP}, \texttt{eht-imaging} and \texttt{SMILI} 
as described in
reference Akiyama K. et al., 2019d, ApJ, 875, L4;
used by the EHT Collaboration for the observation day of April 11 of 2017. 
Images have equivalent resolution. 
}\label{fig:las+tres}
\end{figure*}

\begin{figure*}
\centering
\includegraphics[clip,width=0.45\textwidth]{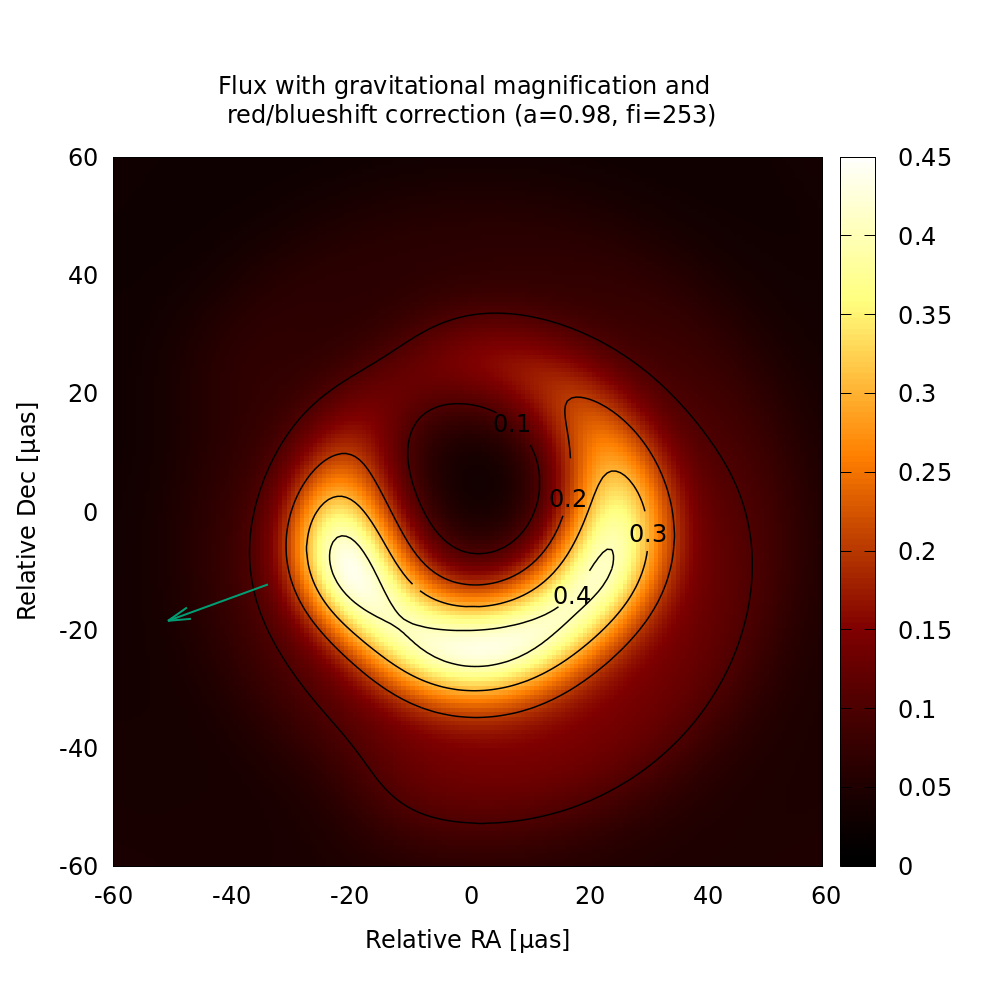}
\includegraphics[clip,width=0.42\textwidth]{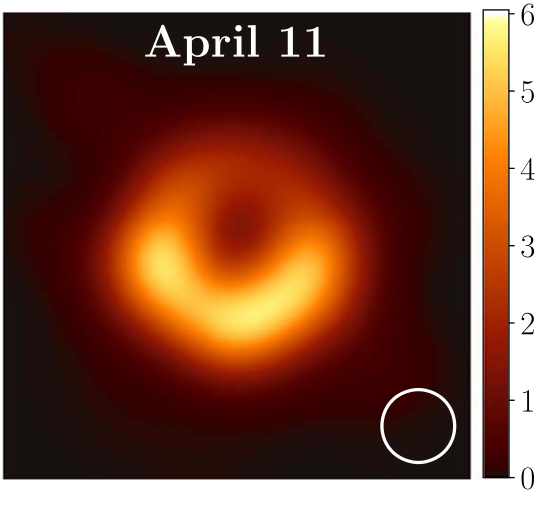}
\caption{
	Comparison between our final image, on the left, with the EHT image of April 11 of 2017,
	on the right.
	Our model is based on a two temperature thin disk 
	with an accreting flow in prograde motion plus an asymmetric bar. 
}\label{fig:final}
\end{figure*}

Our results in Fig. \ref{fig:flux-arrow-5} should be compared with those of the EHT team. 
In particular from reference \cite{Akiyama:2019bqs} we reproduce
the result of three pipelines, appearing in their fig. 14
(see our Fig. \ref{fig:las+tres}).
In Fig. \ref{fig:final} we show on the left, our final image, and on the right
we reproduce the April 11 reconstructed image of the EHT Collaboration for M87,
which appears in their fig. 15.
It should be remarked that the April 11 image is constructed from the fiducial
images shown in Fig. \ref{fig:las+tres}.
Our customized image, shown on the left of Fig. \ref{fig:final},
has great similarities with all tree images in Fig. \ref{fig:las+tres},
and with their final average of the April 11 image (shown on the right of Fig. \ref{fig:final}).
This confirms that such a model, although a very simplified one,
can give account of the present observed features with spectacular fidelity;
since the main qualitative and quantitative astrophysical indicators 
on the EHT image are reproduced.

In our description though, we do not intend to ascribe a clear physical origin 
to a bar like emission model; instead we have just investigated which are 
the appropriate simple geometries that are successful to describe the emitter.

\subsection{Plain quantitative comparison of our images}

Up to this point we have presented our images based on the
aspect that they have to our human eyes; since 
the EHT presentation of the final images intend
to emphasize the human perception.
For example the choice of the color palette was
done in order to emphasize our visual interpretation.
For this reason we have also used the same family
of color palette in our images.
To give a quantitative assessment to image
comparison is a very complicated task, and
it depends strongly on the type of images one
would like to compare.
For instance, there are specific techniques
for dealing with pictures of human faces,
or for dealing with images of fingerprints.
It is our intention here to complement our previous
choices, that conduced us the our final image,
with a simple and elementary quantitative comparison
of our images with the target EHT image of April 11.
The measure we have chosen is the correlation
between two set of data $\mathbf{v}_1$ and $\mathbf{v}_2$,
that we think as vectors in an appropriate vector space,
given by
\begin{equation}\label{key}
\rho = \frac{<\mathbf{v}_1,\mathbf{v}_2>}
{\sqrt{<\mathbf{v}_1,\mathbf{v}_1><\mathbf{v}_2,\mathbf{v}_2>}}
;
\end{equation}
where $<,>$ is the natural scalar product in
the vector space.

\begin{figure}
\centering
\includegraphics[clip,width=0.49\textwidth]{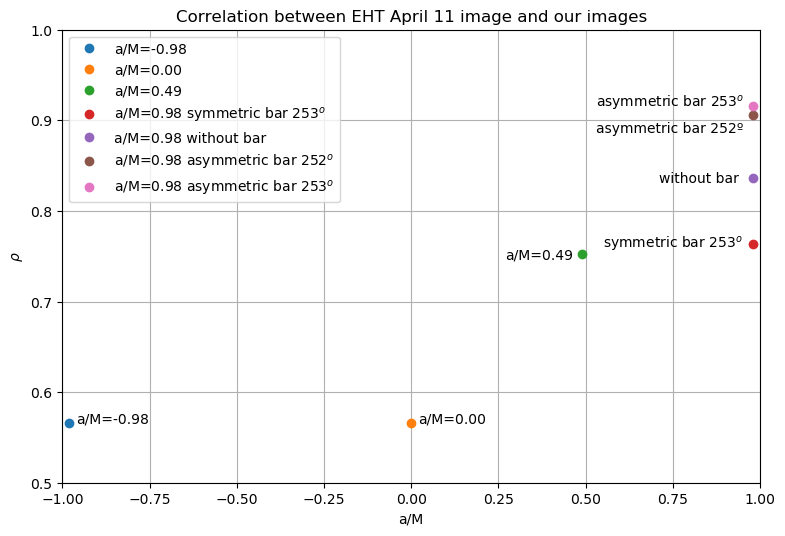}
\caption{
Correlation $\rho$ between the EHT April 11 image with our images
as a function of the angular parameter $a/M$.
For our final choice $a/M=0.98$, we also show
variants with a symmetric emitting bar, without a bar,
and with an asymmetric bar at different angular
orientations. 
The kind of bar and orientation chosen gives
the highest correlation value.
}\label{fig:q-comparisson}
\end{figure}

It should be noted the EHT images, and our images,
are color generated from an intensity
distribution.
For this reason we have first generated back
gray scale images before carrying out
the comparisons.
The standard scalar product is then the
usual Euclidean product of the
pixel values.
The result is presented in Fig. \ref{fig:q-comparisson};
where we show selected values
of the correlation in terms of the
angular parameter $a/M$, with some variants
in the chosen value of $a/M=0.98$.
Although we have made several trials, we
only show a selection, in order not to
overcrowd the graph.
One can see that the asymmetric bar
with the reference orientation at $253^o$
gives the best correlation value;
justifying our selection of the
geometry for the emitting region.

\section{Final comments}\label{sec:Final+commnets}
The recent progress made in the observations of the innermost part of the strong radio 
emissions coming from the center of nearby galaxies by the EHT Collaboration, provides a 
new tool that brings us closer to the possibility of testing our models on very compact 
sources and the surrounding material in their immediate neighborhood.
The study of the physics in these regimes is usually done through the use of
magnetohydrodynamical models.
In particular, for the case of the SMBH in M87, 
the template bank of high resolution images employed in the simulations of the
EHT Collaboration contains as a distinguished characteristic the presence of a
bright emission ring in most of their models; it
seems to indicate that a bright  ring 
could explain the main characteristics of
most of the 
images obtained through the various pipelines at the resolutions allowed by 
the instrumental facilities. 
Such a ringlike feature has been emphasized in \cite{Akiyama:2019cqa} as 
the cause of the crescent 
shape observed in the final reported images such as that reproduced in 
Fig. \ref{fig:EHT+April_11}; and, has been associated with the so-called photon 
ring region present 
in the exterior of Kerr spacetime.
The observable consequences of a photon ring in Kerr spacetime has been discussed 
recently by several authors \citep{Gralla:2019xty, Gralla_2020PhRvD, Johnson_2020} in the 
case of optically thin accretion disk.

Instead, in this work we have approached the simulation process of images from a different 
perspective, without emphasizing the effects of the photon ring model. 
The guiding idea we have employed is that of using the simplest
geometrical model that could account for the main observed features
found in the EHT images. 
We find that
the proper signatures imprinted on the final image should almost be completely 
associated to the emission model and red/blueshift contributions. 
That is, we have not considered the
 enhanced intensity due to the contribution of photons paths with many 
turns around the SMBH. 
This conceptual point becomes clear in our implementation described in section 
\ref{sec:Numerical+implementaion}. 	
We have retained the simplest ingredients in the kinematics and emission
properties that one could demand to the plasma orbiting a Kerr black hole 
based on previous
results in the literature; namely, a major emission taking place on the equatorial plane
with a plasma with two temperature regions orbiting in prograde circular geodesic
motion. 
The geometrically thin disk approximation for the main emission, 
is a common fact in several 
GRMHD models, in particular, this is a quite stable property 
in `magnetically arrested disk' 
\citep{Narayan_2003PASJ}.
Notably, these kind of models are the ones preferred by the new reports on polarization 
from M87\citep{Akiyama:2021tfw}. 
The presence of a two temperature regions is usually ascribed to the lack of thermal equilibrium
between the plasma components compatible also with the general picture for low luminosity active galactic 
nuclei (LLAGN's).
Additionally, the assumption of prograde circular geodesic 
motion seems to support that the kinematic is 
essentially dictated by the geometry on the equatorial plane being the 
red/blueshift factor the main
contribution to the brightenss pattern observed it the images.
This general setting together with a localized bar-like emission at appropriated angle
allows us to describe with impressive similarity the images of M87 generated by the EHT 
Collaboration, as is shown in Fig. \ref{fig:final}.
The only presence of those general basic characteristics and the inclusion of that 
peculiar geometric emission presents an alternative setting to account for the 
observed features in the images constructed from current VLBI observations.

In more detail, in sections \ref{subsec:2T+model} and \ref{subsec:2T+spinvarying} 
the images built from this model give account for the south-west crescent part of the bright 
region observed in Fig. \ref{fig:EHT+April_11}. 
This is achieved for a near extreme Kerr SMBH with $a/M = 0.98$ and a spin direction anti-aligned 
with respect to the large scale jet of M87.
All this is a
robust indication of the relevance of the disk geometry.
But a full account of the whole crescent shape which also shows a bright small 
region at the east 
part is obtained when we add the asymmetric bar like emission feature.
The impressive similarities of our image with the three pipelines and with their  
final image of April 11, can be appreciated from Figs. 
\ref{fig:las+tres} and \ref{fig:final}.
This reveals 
that with the present level of accuracy 
our simple modeling can be employed to give
a very good reconstruction of the image generated from
the observed EHT data.
Although we have concentrated on the April 11 image in the above discussion,
our construction also gives excellent representation of the April 5, 6 and 10
images.

It is worthwhile to mention that our presentation combines
the emission model, that we have described in detail, 
with a novel ray-tracing technique 
that includes the jointly implementation of the null geodesic equations 
and null geodesic deviation 
equations.

In this way we take into account and quantify the whole lenses effects,
by including the contribution of the lens optical scalars,
which is normally neglected in the literature. 

We expect to apply the techniques used in this article
to other astrophysical systems of interest.

\subsection*{Data Availability}
No new data were generated or analysed in support of this research.
The numerical calculation is completely described in the article.

\subsection*{Acknowledgments}

We are grateful to the authors
of the EHT publications, Heino Falcke and Huib Jan van Langevelde,
for the kind interchange we had when we contacted them to ask
permission to reproduce their images. 
We are also grateful to an anonymous Referee for
several criticisms that contributed to the 
improvement of our manuscript.

We acknowledge support from CONICET, SeCyT-UNC and Foncyt.

\appendix

\section*{Appendices}

\section{Christoffel symbols in Boyer-Lindquist coordinate chart}\label{ap:Christ+symb}
Some Christoffel symbols in Boyer-Lindquist coordinates appear in the coupled system 
of equations \eqref{eq:ell+nabla+ell} and \eqref{eq:geod+dev+equation} that we use 
to perform the integration (see appendix \ref{ap:Linear+order+system} below).
Even though in most practical situation one does not require explicit knowledge of them,
it might be of interest to have a reference at hand.
The list is the following:
\begin{equation}
\Gamma^{\;\,t}_{t\;\;r} = \frac{M \left( r^2 + a^2 \right)\big(r^2 - a^2\cos(\theta)^2\big)}
{\Sigma^2 \Delta},
\end{equation}
\begin{equation}
\Gamma^{\;\,t}_{t\;\;\theta} = - \frac{a^2 M r \sin(2\theta)}{\Sigma^2 \Delta},
\end{equation}
\begin{equation}
\Gamma^{\;\,t}_{r\;\;\phi} = - \frac{a M \sin(\theta)^2 
\Big(\left( r^2 + a^2 \right)\big( r^2 - a^2\cos(\theta)^2\big) + 2r^2\Sigma \Big) }
{\Sigma^2 \Delta},
\end{equation}
\begin{equation}
\Gamma^{\;\,t}_{\theta\;\;\phi} = \frac{a^3 M r \sin(2\theta) \sin(\theta)^2}{\Sigma^2},
\end{equation}	
\begin{equation}
\Gamma^{\;\,r}_{t\;\;t} = \frac{ M \Delta \big( r^2 - a^2 \cos(\theta)^2 \big)}{\Sigma^3},
\end{equation}	
\begin{equation}
\Gamma^{\;\,r}_{t\;\;\phi} = - \frac{ a M \Delta \big( r^2 - a^2 \cos(\theta)^2 \big) \sin(\theta)^2}{\Sigma^3},
\end{equation}		
\begin{equation}
\Gamma^{\;\,r}_{r\;\;r} = \frac{ r \Delta + \left( M - r \right)\Sigma }{\Sigma \Delta},
\end{equation}			
\begin{equation}
\Gamma^{\;\,r}_{r\;\;\theta} = - \frac{a^2 \sin(2\theta)}{2\Sigma},
\end{equation}			
\begin{equation}
\Gamma^{\;\,r}_{\theta\;\;\theta} = - \frac{r \Delta}{\Sigma},
\end{equation}			
\begin{equation}
\Gamma^{\;\,r}_{\phi\;\;\phi} = - \frac{\Delta \sin(\theta)^2}{\Sigma^3}
\Big( r\Sigma^2 - a^2 M\left( r^2 - a^2\cos(\theta)^2\right)\sin(\theta)^2 \Big), 
\end{equation}			
\begin{equation}
\Gamma^{\;\,\theta}_{t \;\; t} = - \frac{a^2 M r \sin(2\theta)}{\Sigma^3}, 
\end{equation}				
\begin{equation}
\Gamma^{\;\,\theta}_{t \;\; \phi} = \frac{a M r \left( r^2 + a^2 \right) \sin(2\theta)}{\Sigma^3}, 
\end{equation}				
\begin{equation}
\Gamma^{\;\,\theta}_{r \;\; r} = \frac{a^2 \sin(2\theta)}{2\Sigma \Delta}, 
\end{equation}				
\begin{equation}
\Gamma^{\;\,\theta}_{r \;\; \theta} = \frac{ r}{\Sigma}, 
\end{equation}					
\begin{equation}
\Gamma^{\;\,\theta}_{\theta \;\; \theta} = -\frac{a^2 \sin(2\theta)}{2\Sigma}, 
\end{equation}					
\begin{equation}
\begin{split}
\Gamma^{\;\,\theta}_{\phi \;\; \phi} =& -\frac{\sin(2\theta)}{2\Sigma^3}
\Big( \left(r^2 + a^2\right)\Sigma^2  \\
& \qquad \qquad + 2a^2 Mr \big( \left(r^2 + a^2\right) + \Sigma \, \big)
\sin(\theta)^2 \Big), 
\end{split}
\end{equation}					
\begin{equation}
\Gamma^{\;\,\phi}_{t \;\; r} = \frac{a M \big( r^2 - a^2 \cos(\theta)^2 \big)}{\Sigma^2 \Delta}, 
\end{equation}						
\begin{equation}
\Gamma^{\;\,\phi}_{t \;\; \theta} = -\frac{2 a M r \cos(\theta)}{\Sigma^2 \sin(\theta)}, 
\end{equation}				
\begin{equation}
\Gamma^{\;\,\phi}_{r \;\; \phi} = \frac{r \Sigma \Delta - 
a^2 M \sin(\theta)^2 \big( r\Sigma + r^2 - a^2 \cos(\theta)^2 \big)}
{\Sigma^2 \Delta}, 
\end{equation}								
\begin{equation}
\Gamma^{\;\,\phi}_{\theta \;\; \phi} = \frac{\cos(\theta)}{\sin(\theta)} + 
\frac{a^2 r M \sin(2\theta)}{\Sigma^2}.
\end{equation}

\section{General relation between the surface brightness and specific intensity}\label{ap:B+I}

\subsection{Surface brightness}\label{ap:subsec+SurfBright}
The emission from an extended or diffuse source, is characterized by the 
surface brightness; so that all other quantities related to emission
are determined from this one.

The \emph{surface brightness} quantifies the amount of power radiated
by the projected surface of the source in the direction of sight. 
More precisely, if the surface element $dS_s$ emits a power $\mathscr{P}_s$ in all
directions; the \emph{surface brightness} $B(\theta_s,\phi_s,\nu_s)$ is the amount  
of this power per unit of frequency $\nu_s$, which is emitted per solid angle 
$d\omega_s$ in the direction $(\theta_s, \phi_s)$ and per unit of surface $dA_s$ orthogonal 
to the direction $(\theta_s, \phi_s)$:
\begin{equation}
d\mathscr{P}_s \equiv B(\theta_s,\phi_s,\nu_s) \, d\nu_s \, d\omega_s \, dA_s.
\end{equation}

\subsection{The specific intensity}\label{ap:subsec+Spec+Intens}
Devices used to collect radiation, in general are able to resolve 
extended images within its angle of aperture. 
The radiant power per frequency per unit of collector surface and per solid angle 
is quantified by the \emph{specific intensity} and denoted by $I(\theta_o, \phi_o,\nu_o)$;
it is explicitly defined through:
\begin{equation}
d\mathscr{P}_o \equiv I(\theta_o, \phi_o,\nu_o) \, dA_o \, d\Omega_o \, d\nu_o;
\end{equation}
where $dA_o$ is the surface element of collector orthogonal to the direction
$(\theta_o, \phi_o)$; $d\Omega_o$ denotes the solid angle in the direction of the beam and
$\nu_o$ is the frequency measured by the observer.

\subsection{Photon conservation and the relation between surface brightness and specific intensity}

If the photon number is conserved then, the amount of power radiated $d\mathscr{P}_s$ 
by the source in the direction of the observer only can differ from the 
observed power $d\mathscr{P}_o$ by a redshift factor between the source and the observer:
\begin{equation}
d\mathscr{P}_s = (1 + z)^2 \, d\mathscr{P}_o;
\end{equation}
since holds the relation
\begin{equation}\label{eq:redshift+I}
1 + z = \frac{\nu_s}{\nu_o} = \lim_{\tau_s \to 0}{\frac{\Delta \tau_o}{\Delta \tau_s}}.
\end{equation}

This consideration implies
\begin{equation}
B(\theta_s, \phi_s, \nu_s) dA_s d\omega_s d\nu_s = (1 + z)^2 \, I(\theta_o, \phi_o, \nu_o)
dA_o d\Omega_o d\nu_o;
\end{equation}
or 
\begin{equation}\label{eq:SurfBright-MeasurFlux}
B(\theta_s, \phi_s, \nu_s) dA_s d\omega_s = (1 + z) \, I(\theta_o, \phi_o, \nu_o) dA_o d\Omega_o.
\end{equation}

\subsection{The Etherington theorem}
The Etheringthon theorem is a purely geometric result valid in a general spacetime
relating the set of angles appearing in the definitions of the surface brightness and 
specific intensity.
The theorem states that
\begin{equation}\label{eq:reciprocity-rel}
dA_o d\Omega_o = \left( 1 + z \right)^2 dA_s d\omega_s,
\end{equation}
and for details we refer the reader to the article of \cite{Ellis71}.
From this result it follows immediately that the link between 
surface brightness and specific intensity is:
\begin{equation}\label{eq:B=1+z**3*I}
B(\theta_s, \phi_s, \nu_s) = (1 + z)^3 I(\theta_o, \phi_o, \nu_o).
\end{equation}
Here it is important to remark that this relation is independent of the distance; 
it only depends on redshift.

\subsection{Equivalence	between the intensity magnification and the angular magnification}
In the study of gravitational lenses, it is at the core of most 
analysis the fact that magnification measured in terms of cross section
of thin bundles and magnification measured in terms of collected intensity
coincide.
In reference \cite{Boero:2016nrd} we have defined the 
\emph{intensity magnification} for unresolved sources as the quotient
of the following quotient of fluxes
\begin{equation}
\tilde{\mu}(\lambda) \equiv \frac{\mathscr{F}(\lambda,z)}{\mathscr{F}_0(\lambda,z)},
\end{equation}
where $\mathscr{F}(\lambda,z)$ denotes the observed flux of an unresolved
object at affine distance $\lambda$ and relative motion determined by the redshift
$z$, in a general space–time, while $\mathscr{F}_0(\lambda,z)$ denotes the flux that
one would expect to collect from the same object at the same distance $\lambda$
in Minkowski space–time with the same relative motion.

For the case of extended (i.e. resolved) sources the most natural observable
is the specific intensity, and therefore we define the 
\emph{intensity magnification} in these case in terms of quotients of observed 
specific intensities.

The observed flux received per unit of frequency, from a source at affine 
distance $\lambda$ is given by $I(\theta_o, \phi_o, \nu_o, \lambda)d\Omega_o$
while the flux that one would expect from the the same source at the same 
distance in Minkowski spacetime and with relative velocity given
by the observed redshift $z$ is 
$I_0(\theta_o, \phi_o, \nu_o, \lambda, z) d\Omega_{0_o}$; we then 
define the \emph{intensity magnification} as follows:
\begin{equation}\label{eq:mu+tilde=mu+resolved}
\tilde{\mu}(\lambda) \equiv \frac{I(\theta_o, \phi_o, \nu_o, \lambda)d\Omega_o}
{I_0(\theta_o, \phi_o, \nu_o, \lambda, z) d\Omega_{0_o}}.
\end{equation}
Since we are assuming the same redshift conditions, the specific intensities
in the numerator and denominator are the same due to equation
\eqref{eq:B=1+z**3*I}; and therefore we obtain again the 
equivalence between the intensity magnification and the angular
magnification, this is:
\begin{equation}
\tilde{\mu} = \mu \equiv \frac{d\Omega_o}{d\Omega_{0_o}}.
\end{equation}

Given a model for the surface brightness $B(\theta_s, \phi_s, \nu_s)$ of the source, 
then the observed flux within each pixel of an image is given by
\begin{equation}
\begin{split}
I(\theta_o, \phi_o, \nu_o, \lambda)d\Omega_o \delta \nu_o =& 
\frac{B(\theta_s, \phi_s, \nu_s) d\Omega_o \delta \nu_s}{\left(1 + z \right)^4}
\\
=& 
B(\theta_s, \phi_s, \nu_s)d\Omega_{0_o} \frac{d\Omega_o}{d\Omega_{0_o}}
\frac{\delta \nu_s}{\left(1 + z \right)^4}
\\
=&
B(\theta_s, \phi_s, \nu_s)d\Omega_{0_o} 
\frac{\mu  \, \delta \nu_s}{\left(1 + z \right)^4}
.
\end{split}
\end{equation}

In terms of fluxes the above expression is simply equation \eqref{eq:flux+frequency}:
\begin{equation}
\mathscr{F} \delta \nu_o = \frac{ \mu}{\left(1 + z \right)^4} \mathscr{F}_0 \delta \nu_s 
. \tag{\ref{eq:flux+frequency}}
\end{equation}

\section{Timelike circular orbits around black holes}\label{ap:Circular+orbits}
For completeness, in this appendix we recall the different radius that
characterize the timelike circular geodesic motions in the equatorial plane of a Kerr black hole.

It is customary to address the discussion of timelike circular 
geodesics on the equatorial plane 
in terms of the  prograde (or direct) orbits and 
retrograde orbits separately.

\subsection{Prograde orbits:}
For direct orbits,  
circular geodesics do not exist for radius lesser than 
\begin{equation}
 r_c = 2M\left[ 1 + \cos\left( \frac{2}{3} \arccos\left(-\frac{a}{M}\right) \right)\right];
\end{equation}
unstable orbits take place in the range 
$r_c \leq r \leq r_{\text{ISCO}}$, where
\begin{equation}
r_{\text{ISCO}} = 
M \bigg(
3 + z_2 - \sqrt{(3 - z_1)(3 + z_1 + 2 z_2)}
\bigg);
\end{equation}
with
\begin{align}
z_1 =& 
1 + \sqrt[3]{ 1 - \frac{a^2}{M^2} } 
\bigg( \sqrt[3]{ 1 + \frac{a^2}{M^2} } + \sqrt[3]{ 1 - \frac{a^2}{M^2} }\bigg),
\label{eq:z1}
\\
z_2 =& \sqrt{3 \frac{a^2}{M^2} + z_1^2 }
\label{eq:z2}
;
\end{align}
while stable circular orbits can be found in the region $r_{\text{ISCO}} < r$.
Here 
the label ISCO in the subindex means innermost stable circular orbit. Let us note that in
\cite{Bardeen:1972fi} they are called marginally stable radii and denoted by $r_{\text{ms}}$.
In the discussion of circular orbits it is also worthwhile to mention that even though circular orbits 
are limited to compact regions, there is a subset of them that 
have energies $E_{e} \geq 1$
and therefore, if we perturb them appropriately they could escape to infinity.
In reference \cite{Bardeen:1972fi} they are referred as unbounded circular orbits.
Then, it is useful to look at the range interval at which they take place; following to 
\cite{Bardeen:1972fi} one can find $r_{c} < r < r_{\text{mb}}$ where
\begin{equation}
r_{\text{mb}} = 2M - a + 2\sqrt{M}\sqrt{M - a}.
\end{equation}

\subsection{Retrograde orbits:}
For retrograde orbits,  
circular geodesics do not exist for radius lesser than 
\begin{equation}
 r_c = 2M\left[ 1 + \cos\left( \frac{2}{3} \arccos\left(\frac{a}{M}\right) \right)\right];
\end{equation}
unstable orbits take place in the range 
$r_c \leq r \leq r_{\text{ISCO}}$, where
\begin{equation}
r_{\text{ISCO}} = 
M \bigg(
3 + z_2 + \sqrt{(3 - z_1)(3 + z_1 + 2 z_2)}
\bigg);
\end{equation}
with $z_1$ and $z_2$ defined as in equations \eqref{eq:z1} and \eqref{eq:z2}.
Unstable circular orbits with energies $E_e \geq 1$ now take place in the range 
	$r_c < r < r_{\text{mb}}$, where
\begin{equation}
r_{\text{mb}} = 2M + a + 2\sqrt{M}\sqrt{M + a}.
\end{equation}

\begin{figure}
\centering
\includegraphics[clip,width=0.48\textwidth]{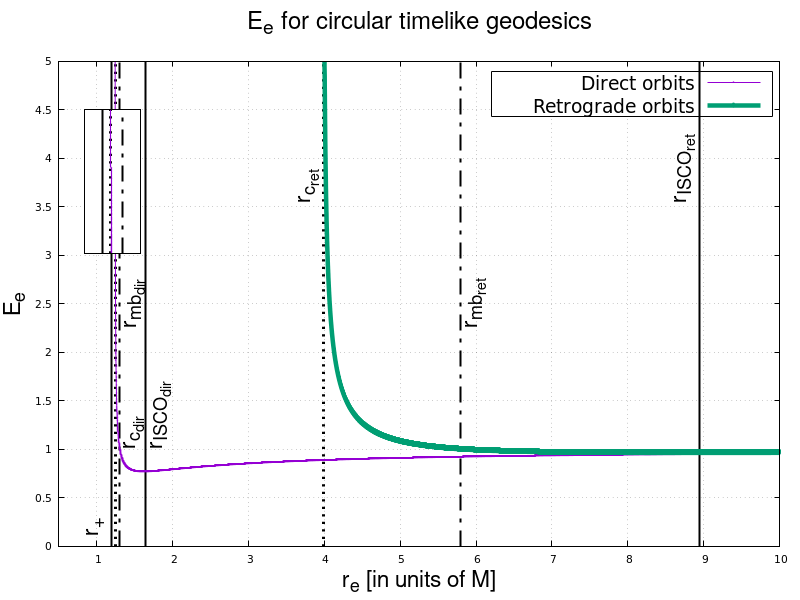}
\caption{Energy for the timelike circular geodesics on the equatorial plane 
$\theta_e = \frac{\pi}{2}$ for a spinning black hole with $a/M = 0.98$.
Green thick curve corresponds to retrograde orbits while thin violet curve
to prograde orbits. 
Continuous vertical lines signal the radius of the event horizon $r_+$
and the radius of the innermost stable circular orbits (ISCO) in both cases 
prograde and retrograde. Doted vertical lines signal the radius of the 
unstable circular photon orbits $r_{c_{\text{dir}}}$ and 
$r_{c_{\text{ret}}}$ for the direct and retrograde cases respectively.
Dash-doted vertical lines signal the radii $r_{mb_{\text{dir}}}$ and 
$r_{mb_{\text{ret}}}$ for the direct and retrograde cases respectively. 
     }
\label{fig:En+timelike+orbits}
\end{figure}

\bsp	
\label{lastpage}
\end{document}